\documentclass[review]{mcom-l}%

\usepackage{amsmath}
\usepackage{amsfonts}
\usepackage{amssymb}
\usepackage{amsthm}
\usepackage{bm}
\usepackage{indentfirst}
\usepackage{graphicx}
\usepackage{subfigure}
\usepackage{array}
\usepackage{fullpage}

\DeclareMathAlphabet{\mathsfsl}{OT1}{cmss}{m}{sl}

\newcommand{\PreserveBackslash}[1]{\let\temp=\\#1\let\\=\temp}
\newcolumntype{C}[1]{>{\PreserveBackslash\centering}p{#1}}
\newcolumntype{R}[1]{>{\PreserveBackslash\raggedleft}p{#1}}
\newcolumntype{L}[1]{>{\PreserveBackslash\raggedright}p{#1}}

\numberwithin{equation}{section}

\theoremstyle{definition}

\usepackage{graphicx} 

\usepackage[utf8]{inputenc}
\usepackage{bm}
\usepackage{mathrsfs}
\usepackage{amsmath}
\usepackage{amsfonts}
\usepackage{amsthm}
\usepackage[ruled]{algorithm} 
\usepackage[noend]{algpseudocode} 
\usepackage{color}
\usepackage{setspace}
\usepackage{float}
\usepackage{import}
\usepackage{url}
\usepackage{multicol}
\usepackage{multirow}
\usepackage{subfigure}
\usepackage[sort&compress,numbers]{natbib}
\usepackage[top=1in,bottom=1in,left=1in,right=1in]{geometry}
\usepackage{hyperref}
\usepackage[misc]{ifsym}
\usepackage{tikz}

\newcommand{\real}{\mathbb{R}}

\newcommand{\mcD}{\mathcal{D}}

\newcommand{\mcG}{\mathcal{G}}

\newcommand{\mcJ}{\mathcal{J}}

\usepackage{xcolor}

\def\omg{{\Omega}}

\def \bb{\bm{b}}

\def \fb{\bm{f}}

\def \Ib{\bm{I}}
\def \Fb{\bm{F}}

\def \ub{\bm{u}}
\def \Ub{\bm{U}}
\def \sigmab{\bm{\sigma}}
\def \Eb{\bm{E}}
\def \wb{\bm{w}}
\def \vb{\bm{v}}
\def \xb{\bm{x}}

\def \hb{\bm{h}}

\def \Kb{\bm{K}}

\def \qb{\bm{q}}

\def \yb{\bm{y}}
\def \cb{\bm{c}}
\def \Cb{\bm{C}}

\def \xib{{\boldsymbol\xi}}
\def \etab{{\boldsymbol\eta}}

\def \Wb{\bm{W}}

\def \mbF{\mathbb{F}}
\def \mbU{\mathbb{U}}

\newcommand{\ue}{{\underline e}}
\newcommand{\ubM}{{\underline{\bm M}}}
\newcommand{\ubT}{{\underline{\bm T}}}

\newcommand{\ut}{{\underline t}}
\newcommand{\ubY}{{\underline{\bm Y}}}
\newcommand{\Rb}{{\bm R}}

\newcommand{\vertii}[1]{{\left\vert\left\vert #1
    \right\vert\right\vert}}    
    
\newcommand{\verti}[1]{{\left\vert #1
    \right\vert}}

\newcommand{\YY}[1]{{\color{black}#1}}

\title{Heterogeneous Peridynamic Neural Operators: Discover Biotissue Constitutive Law and Microstructure From Digital Image Correlation Measurements}

\author[S. Jafarzadeh, S. Silling, L. Zhang, C. Ross, C-H. Lee, R. Rahman, S. Wang, Y. Yu]{}

\thanks{S. Jafarzadeh would like to acknowledge support by the AFOSR grant FA9550-22-1-0197, and Y. Yu would like to acknowledge support by the National Science Foundation under award DMS-1753031. Portions of this research were conducted on Lehigh University's Research Computing infrastructure partially supported by NSF Award 2019035.} 
\thanks{C.-H. Lee would like to acknowledge the support from the National Institute of Health (Grant No. R01HL159475-01A1).}

\thanks{This article has been authored by an employee of National Technology and Engineering Solutions of Sandia, LLC under Contract No. DE-NA0003525 with the U.S. Department of Energy (DOE). The employee owns all right, title and interest in and to the article and is solely responsible for its contents. The United States Government retains and the publisher, by accepting the article for publication, acknowledges that the United States Government retains a non-exclusive, paid-up, irrevocable, world-wide license to publish or reproduce the published form of this article or allow others to do so, for United States Government purposes. The DOE will provide public access to these results of federally sponsored research in accordance with the DOE Public Access Plan https://www.energy.gov/downloads/doe-public-access-plan.}

\thanks{$^*$Corresponding author: Yue Yu (\href{yuy214@lehigh.edu}{yuy214@lehigh.edu})}

\begin{document}

\maketitle

\centerline{\scshape
Siavash Jafarzadeh$^{1}$,
Stewart Silling$^{2}$,
Lu Zhang$^{1}$,
Colton Ross$^{3}$,}

\centerline{\scshape
Chung-Hao Lee$^{3,4}$,
S. M. Rakibur Rahman$^{5}$,
Shuodao Wang$^{5}$,
and Yue Yu$^{{\href{mailto:yuy214@lehigh.edu}{\textrm{\Letter}}}*1}$}

\medskip

{\footnotesize
 \centerline{$^1$Department of Mathematics, Lehigh University, Bethlehem, PA 18015, USA}
} 

\medskip

{\footnotesize
 \centerline{$^2$Center for Computing Research, Sandia National Laboratories, Albuquerque, NM, USA}
}

\medskip

{\footnotesize
 \centerline{$^3$School of Aerospace and Mechanical Engineering, The University of Oklahoma, Norman, OK 73019, USA}
}

\medskip

{\footnotesize
 \centerline{$^4$Department of Bioengineering, University of California, Riverside, Riverside, CA 92521, USA}
}

\medskip

{\footnotesize
 \centerline{$^5$School of Mechanical and Aerospace Engineering, Oklahoma State University, Stillwater, OK 74078, USA}
}

\bigskip





\begin{abstract}
Human tissues are highly organized structures with collagen fiber arrangements varying from point to point. Anisotropy of the tissue arises from the natural orientation of the fibers, resulting in location-dependent anisotropy. Heterogeneity also plays an important role in tissue function. It is therefore critical to discover and understand the distribution of fiber orientations from experimental mechanical measurements such as digital image correlation (DIC) data. To this end, we introduce the Heterogeneous Peridynamic Neural Operator (HeteroPNO) approach for data-driven constitutive modeling of heterogeneous anisotropic materials. Our goal is to learn a nonlocal constitutive law together with the material microstructure, in the form of a heterogeneous fiber orientation field, from load-displacement field measurements. We propose a two-phase learning approach. First, we learn a homogeneous constitutive law in the form of a neural network-based kernel function and a nonlocal bond force.
This captures the complex homogeneous material responses from data. Second, we reinitialize the learnt bond force and train it together with the kernel and the fiber orientation field for each material point. Due to the inherent properties of the state-based peridynamic framework that we use, our HeteroPNO-learned material models are guaranteed to be objective and to satisfy the balances of linear and angular momenta. The effects of heterogeneity and the nonlinearity of the constitutive relationship are captured by the nonlocal kernel function, enabling physical interpretability. These features enable our HeteroPNO architecture to learn a constitutive model for biological tissue with anisotropic, heterogeneous response undergoing large deformation.  To demonstrate the applicability of our approach, we apply the HeteroPNO in learning a material model and fiber orientation field from DIC measurements in biaxial testing. We evaluate the learnt fiber architecture, which is found to be  consistent with observations from polarized spatial frequency domain imaging. The framework is capable of providing displacement and stress field predictions for new and unseen loading instances.
\end{abstract}

\maketitle

\clearpage
\tableofcontents

\section{Introduction}


Biological tissues are highly organized structures with specific architectures that enable local functions. Tissues exhibit a unique ability to adapt, remodel, and self-heal in cases of disease or injury \cite{kuhl2005remodeling,ambrosi2011perspectives,hinz2012recent,irons2020cell,howsmon2021valve,irons2021transcript}. Many tissue structures are heterogeneous with specific collagen matrix arrangements varying from point to point within the same tissue, allowing it to respond to diverse local mechanical stimuli. Signaling pathways and chemical transduction processes in the underlying cells detect these stimuli, alter the composition and organization of tissue extracellular matrix (ECM), and orchestrate changes in tissue function in response \cite{johnson2021parameterization,aupperle2012pathology,chen2016microstructure,humphrey2000structure,laurence2020pilot,sacks2019simulation}.  Heterogeneity plays an important role in tissue function and resilience.

For the purpose of enabling virtual screening to identify and optimize therapeutic interventions, much effort has been devoted to describing the complex fiber-matrix interactions in the bio-tissues. 
A classical computational approach is based on the development of constitutive relations that are needed for the governing equations.  
In \cite{fung1979pseudoelasticity}, seminal phenomenological constitutive models were employed for the modeling of soft tissues, including the iris \cite{pant2018appropriate}, cardiac heart valves \cite{may1998constitutive,prot2007transversely,sacks2016novel}, arterial vessels \cite{van2011generic}, and skin \cite{bischoff2000finite}. In \cite{lee2014inverse,johnson2021parameterization,kamensky2018contact}, the authors developed a transversely isotropic model through an inverse modeling approach. They employed it to characterize the in-vivo mechanical response of the mitral valve anterior leaflet. To model the adaptations of fiber architecture under stress, an ensemble fiber stress–strain relationship was proposed in \cite{fan2014simulation,lee2015effects} and employed to study fiber reorientation and fiber recruitment as observed in experiments. In all of these studies, a strain energy density function is predefined with a specific functional form when modeling the mechanical responses from experimental measurements. Within this predefined material model, the material parameters are calibrated through an inverse method or analytical stress--strain fitting.

\YY{Generally, the use of a predefined material model with the standard continuum mechanics equations allows us to incorporate physics knowledge, such as material objectivity and the balance of linear and angular momentum. Therefore, when the underlying physical mechanism is fully known, such a model could provide a physically admissible description of the material response, even when only a small amount of observable data is available for calibration \cite{movahhedi2023predicting,wu2024identifying}. However, with many biological tissues, an understanding of their underlying physical mechanism remains limited, potentially diminishing the reliability of a predefined model. There are two reasons for this.} First, the descriptive power of the models may be restricted to certain deformation modes and strain ranges, thus restricting its predictivity and generalizability \cite{lee2014inverse,he2021manifold,lee2017vivo}.  Also, for some tissues such as skin \cite{limbert2019skin,tacc2023benchmarking}, there is currently no definitive material model available in the literature that would provide the predefined model. Second, many of these models are developed for \emph{homogeneous} materials, and hence a full description that accounts for the microstructure is not provided by global stress-strain measurements.  \YY{Although pre-defined random fields can be introduced to further provide a stochastic modeling of nonlinear elasticity \cite{tacc2024generative}, in reality, they cannot fully characterize the mechanical properties and microstructure of bio-tissues because these parameters not only vary between tissue types, but also vary with position and can be different for different patients.} As a result, there is a need for a modeling approach that captures the mechanical response of the particular microstructure within a given sample.

To help overcome these challenges, data-driven computing has been considered in recent years as an alternative. In the context of biological tissue modeling  \cite{pfeiffer2019learning,he2021manifold,he2021deep,tac2021data}, data-driven constitutive laws directly integrate material identification with the modeling procedures, and hence do not require a predefined constitutive model form. 
In \cite{minano2018wypiwyg}, the constitutive law for soft tissue damage was constructed by solving the system of linear equations consisting of coefficients of shape functions, rather than nonlinear fitting to a predefined model.  A local convexity data-driven (LCDD) computational framework was developed in \cite{he2021manifold,he2021deep}. This framework couples manifold learning with nonlinear elasticity and was applied to model the stress--strain response of the porcine mitral (heart) valve posterior leaflet. In \cite{tac2021data}, a neural network was developed to infer the relationship between the isochoric strain invariants and the value of strain energy. This was applied to learn the mechanical behavior of porcine and murine skin from biaxial testing data. Despite these advances, data-driven constitutive laws on soft tissue modeling have mostly focused on the identification of stress--strain and energy--strain relationships for a homogenized material model, and are thus unable to capture the effects of heterogeneity. 
Because of the importance of heterogeneity in tissue function and failure \cite{howell2017role,you2022physics}, the assumption of homogeneity seriously restricts the capabilities of a constitutive model.

In an alternative approach, there has been significant progress in the development of deep neural networks (NNs) for heterogeneous material modeling \cite{wang2018multiscale,he2020physics,tartakovsky2020physics,liu2019deep,yang2019derivation,garbrecht2021interpretable,you2022learning,lu2019deeponet,lu2021learning,li2020neural,li2020multipole,li2020fourier}. Among these works, we focus on the neural operator learning approach \cite{you2022learning,lu2019deeponet,lu2021learning,li2020neural,li2020multipole,li2020fourier,liu2024domain,yin2024dimon}, which learns the maps between the function spaces. In comparison with classical NNs, the most notable advantage of neural operators is their generalizability to different input instances. As a result, measurements with different resolutions can be integrated to train the same model \cite{you2022data}, and the trained model can be used to solve for new instances on a different grid \cite{you2022nonlocal}. Moreover, the neural operator is provably a universal approximator for function-to-function mappings \cite{chen1996reproducing,lanthaler2023nonlocal}. Therefore, it possesses the capability to capture both complex material responses and the heterogeneity within data. All of these properties make the neural operator particularly promising for learning complex material responses without predefined constitutive models and direct measurements of the microstructure. In the context of continuum mechanics, most of the current neural operators have assumed that the underlying mathematical description uses partial differential equations (PDEs). The neural operator learns a surrogate mapping from the loading field to the material response field, which can be seen as learning the solution operator of a hidden PDE. In \cite{yin2022simulating,goswami2022physics,yin2022interfacing}, neural operators have been successfully applied to modeling the physical response of homogeneous materials. In \cite{li2020neural,li2020multipole,li2020fourier,lu2021comprehensive}, neural operators were used as a solution surrogate for Darcy flow in a heterogeneous porous medium with a known microstructure field. In \cite{you2022learning}, an implicit neural operator architecture, namely the implicit Fourier neural operator (IFNO), was proposed to model material responses without using any predefined constitutive models or microstructure measurements. IFNOs were then applied to learn the tissue responses from digital image correlation (DIC) measurements. It was found that this neural operator approach is superior to other methods at capturing the heterogeneous features \cite{you2022physics}.

Despite these advances in neural solution operators, this type of methodology has three fundamental limitations. Firstly, it is challenging to incorporate fundamental physical laws into the solution operators. As a result, the solution operator needs to learn all physical laws from data, and therefore it cannot always guarantee the physical consistency of fundamental laws such as conservation of linear and angular momentum and Galilean and frame invariances. Secondly, since the solution operator varies under different domain shapes and boundary conditions, the neural operators are not generalizable to different domain geometries and loading scenarios. That means, for example, an operator learned on square-domain samples cannot be immediately applied to predict the material deformation on a circular domain. Finally, with solution operators, the information about the constitutive law and microstructure is captured implicitly via neural network parameters. Therefore, the complex relation between the constitutive law and the microstructure is largely hidden, making it challenging to interpret the learnt model. To overcome the first two challenges, in \cite{jafarzadeh2024peridynamic} we proposed the Peridynamic Neural Operator model (PNO), which provides a homogeneous nonlocal constitutive model. This neural operator provides a forward model in the form of state-based peridynamics, with Galilean invariance, frame invariance, and momentum balance laws guaranteed. The model was validated on a DIC displacement tracking dataset, and it was compared to baseline models that use predefined constitutive laws. This PNO achieved improved accuracy by learning the constitutive law from data, without the need to learn the required balance laws and invariances. It also showed generalizability to different domain configurations, external loadings, and discretizations.

The PNO described in \cite{jafarzadeh2024peridynamic} assumed homogeneity. In the present work, we take this approach one step further and develop a data-driven constitutive law for heterogeneous materials. \YY{While there is some earlier  work on learning spatially varying functions from noisy data \cite{biegler2011large, biegler2003large, antil2018frontiers, ghattas2021learning, stuart2010inverse}, to the best of our knowledge, learning a heterogeneous constitutive model based on neural networks for materials with spatially varying
properties has not been studied before in the literature}. Specifically, we propose the Heterogeneous Peridynamic Neural Operator (HeteroPNO) and apply it to learn the material model together with fiber orientation field from DIC measurements. We apply this to data for a representative tricuspid valve anterior leaflet (TVAL) specimen from a porcine heart. \YY{Such an extension from learning homogeneous models to heterogeneous ones is non-trivial. From the PDE-based modeling standpoint, inferring the underlying microstructure is an inverse problem, which is generally an enduring ill-posed problem especially when the measurements are scarce \cite{fan2023solving,molinaro2023neural,jiang2022reinforced,chen2023let}. Unfortunately, the ill-posedness issue may become even more severe in neural network models, due to the inherent bias of neural network approximations \cite{xu2019frequency}. Inspired by the regularization methods of inverse PDE problems which incorporate prior information \cite{dittmer2020regularization,obmann2020deep,ding2022coupling,chen2023let}, we propose in the present work to address this challenge by imposing physical knowledge, which can be seen as an analog to prior information, in model design. In particular, we design HeteroPNO such that the fiber orientation is captured through an anisotropic position-dependent nonlocal kernel, and material nonlinearity is represented through a nonlinear dependence of the internal forces on the deformation.} As a result, the effects of the material heterogeneity and the constitutive law are disentangled. The position dependence of the collagen fiber orientation are captured by rotating the kernel to align its principal direction with that of the fibers. Based on this architecture, we develop a two-phase learning algorithm. First, we learn a homogeneous PNO model to capture the kernel form and the constitutive law. Second, we fine-tune the model by learning a fiber orientation field as a function of position. To evaluate the performance of the HeteroPNO, we validate the discovered collagen fiber architecture by comparing it with measurements from polarized spatial frequency domain imaging (pSFDI)-biaxial testing system \cite{fitzpatrick2022ex}. We assess the predictability of our model reproducing displacement and stress fields under unseen scenarios. To the best of our knowledge, the present work is the first time that a neural operator learning approach has been applied to discover both the constitutive law and microstructure from data.

The remainder of this paper is organized as follows. In Section \ref{sec:background}, we first review the background of the peridynamic nonlocal mechanical theory, the nonlocal neural operators, and the homogeneous Peridynamic Neural Operator (PNO). Then, our proposed architecture, which is based on PNO and its incorporation of heterogeneity through a pointwise fiber orientation field, is introduced in Section \ref{sec:heteroPNO}. This includes the proposed two-phase machine learning algorithm to discover both the model and microstructure. To verify the performance of our model, in Section \ref{sec:HGO} we demonstrate the effectiveness of the learning technique on a synthetic dataset describing the deformation of a hyperelastic and anisotropic fiber-reinforced material. This demonstration shows the capability of the method in learning the heterogeneous fiber orientation field and reproducing displacement and stress fields consistent with ground-truth solutions. In Section \ref{sec:tissue}, we combine the method with DIC experimental measurements on a representative TVAL specimen. For comparison with alternative methods, we evaluate the results of the proposed HeteroPNO against modeling results obtained using a fitted Fung-type homogeneous constitutive model. Finally, we provide a summary of our key findings and concluding remarks in Section \ref{sec:conclusion}.

\section{Background}\label{sec:background}

In this section, we briefly summarize the peridynamic theory and nonlocal neural operators, including our starting point, which is the Peridynamic Neural Operator for homogenized materials \cite{jafarzadeh2024peridynamic}. Throughout this paper, we use unbolded case letters to denote scalars/scalar-valued functions, bold letters to denote vectors/vector-valued functions, underlined unbolded letters for scalar-valued peridynamic state functions, underlined bold letters for vector-valued state functions, and calligraphic letters for operators.
For any vector $\vb$, we use $\verti{\vb}$ to denote its $l^2$norm. For any function $\fb(\xb)$, $\xb\in\omg \subseteq\real^d$, taking values at nodes $\chi :=\{\xb_1,\xb_2,\dots,\xb_M\}$, $\vertii{\fb}_{l^2(\omg)}$ denotes its $l^2(\omg)$ norm, i.e., $\vertii{\fb}_{l^2(\omg)}:=\sqrt{\sum_{i=1}^M {\verti{\fb(\xb_i)}^2}/M}$, which can be seen as an approximation to the $L^2(\omg)$ norm of $\fb\YY{(\xb)}$ (up to a constant) \YY{assuming uniform grid-spacing}. Here, $\real^d$ represents the dimension-$d$ Euclidean space.


\subsection{Peridynamic Theory}\label{subsec-peri}

Peridynamics provides a description of continuum mechanics in terms of integral operators rather than classical differential operators \cite{silling2000reformulation,seleson2009peridynamics,parks2008implementing,zimmermann2005continuum,emmrich2007analysis,du2011mathematical,bobaru2016handbook}. These nonlocal models include a length scale $\delta$, referred to as the {\emph{horizon}}, which denotes the extent of nonlocal interaction. Because a peridynamic model does not require smoothness of the deformation, it allows a natural description of processes requiring reduced regularity in the solution, such as fracture \cite{du2013nonlocal,yu2021asymptotically,fan2022meshfree}. 

In peridynamics, considering a domain of interest, $\omg\subset\real^d$, the equation of motion is given 
in terms of the displacement $\ub$ as follows:
\begin{equation}
    \rho(\xb)\ddot\ub(\xb,t)= \int_{{B_\delta(\xb)}}\fb(\ub,\qb,\xb,t)\;d\qb+\bb(\xb,t)\text{ ,}\quad (\xb,t)\in\omg\times[0,T]\text{ ,}
\label{eqn-pdeomy}
\end{equation}
where $\xb$ and $\qb$ are material points in the reference (undeformed) configuration of the body. 
$\rho(\xb)$ is the mass density function.
$B_\delta(\xb)$ is a ball centered at $\xb$ of radius $\delta$. 
$\bb(\xb,t)$ is the body force density (external loading), which is assumed to be prescribed.
$\fb(\ub,\qb,\xb,t)$ is the {\emph{pairwise bond force density}} that $\qb$ exerts on $\xb$, satisfying
$\fb(\ub,\qb,\xb,t)=-\fb(\ub,\xb,\qb,t).$
The pairwise bond force density is given by
\begin{equation}
  \fb(\ub,\qb,\xb,t)=
  \ubT[\ub,\xb,t]\langle\qb-\xb\rangle-\ubT[\ub,\qb,t]\langle\xb-\qb\rangle\text{ ,}
\label{eqn-Tdef}
\end{equation}
where the underlined symbols denote {\emph{states}}. States are mappings from a bond $\qb-\xb$ to some other quantity, usually either a vector or a scalar. \YY{In \eqref{eqn-Tdef},} $\ubT$ is called the {\emph{force state}}, which contains the contribution of the material model at a point to the bond force density. The quantities square brackets, $[\ub,\xb,t]$, indicate that $\ubT$ is defined at material point $\xb$ and time $t$, and it is dependent on the displacement field $\ub(\cdot,\cdot)$. The force states in the right hand side of \eqref{eqn-Tdef} contain the contributions of the material models at both bond endpoints $\xb$ and $\qb$ to the pairwise bond force density. 
\YY{
The global balance of linear momentum is necessarily satisfied by the equation of motion
\eqref{eqn-pdeomy} in which the pairwise force density is of the form \eqref{eqn-Tdef}.
The proof of this is given in Proposition 7.1 in \cite{silling2007peridynamic} and follows
by simply computing the rate of change of the total momentum, with the result
\begin{equation}
    \int_\Omega \big(\rho\ddot\ub(\xb, t)-\bb(\xb, t)\big)\,d\xb = {\bf 0}.
    \label{eqn-globalmomentum}
\end{equation}
}A {\emph{material model}} $\hat\ubT(\ubY)$ provides the force state $\ubT$ as a function of the {\emph{deformation state}} $\ubY$, which is defined by
\begin{equation}
   \ubY[\ub,\xb,t]\langle\qb-\xb\rangle=\xib+\etab, \text{ where }\xib:=\qb-\xb,\;\etab:=\ub(\qb,t)-\ub(\xb,t)\text{ .}
\label{eqn-Ydef}
\end{equation}
Thus, in the peridynamic setting, a material model is a state-valued function of a state, rather than a tensor-valued function of a tensor, as in the conventional theory of continuum mechanics.
For a heterogeneous body, our goal is to determine the material model:
\[
  \ubT[\ub,\xb,t]=\hat\ubT(\ubY[\ub,\xb,t],\xb),
\]
where the relationship between the force state and the deformation state depends explicitly on the position $\xb$ as well as the deformation state at that position, $\ubY[\ub,\xb,t]$.

Following \cite{jafarzadeh2024peridynamic}, in this work we assume that the material model is {\emph{ordinary}}, meaning that the bond force vectors in the force state are always parallel to the deformed bonds. As in \cite{jafarzadeh2024peridynamic}, it is further assumed that the material model is {\emph{mobile}}, which means that the magnitudes of the bond force vectors in $\ubT$ depend only on the length changes of the bonds. Denoting the unit direction of the deformed bond as:
\begin{equation}
   \ubM[\ub,\xb,t]\langle\qb-\xb\rangle:=\dfrac{\ubY[\ub,\xb,t]\langle\qb-\xb\rangle}{\verti{\ubY[\ub,\xb,t]\langle\qb-\xb\rangle}}=\dfrac{\xib+\etab}{\verti{\xib+\etab}},
\label{eqn-Mdef}
\end{equation}
and the length changes of the bond as
\begin{equation}
   \ue[\ub,\xb,t]\langle\qb-\xb\rangle:=\verti{\xib+\etab}-\verti{\xib},
\label{eqn-edef}
\end{equation}
the material model for a heterogeneous body composed of ordinary, mobile material can be written as:
\begin{equation}
 \hat\ubT(\ubY,\xb) = \ut(\ue,\xb)\,\ubM\text{ ,}
\label{eqn-mobile}
\end{equation}
\YY{where $\ut$ denotes the scalar force state.} As discussed in \cite{silling2007peridynamic,jafarzadeh2024peridynamic}, this formulation guarantees linear and angular momentum conservation, Galilean invariance, and frame invariance (objectivity).
An important feature of \eqref{eqn-mobile} is that the force in a given bond $\qb-\xb$ can depend on the length changes in {\emph{all}} the bonds in the family of $\xb$; expressing this dependence precisely is the purpose of state-based peridynamic material models.

Combining \eqref{eqn-pdeomy}, \eqref{eqn-Tdef}, and \eqref{eqn-mobile}, we obtain the following peridynamic model: 
\begin{multline}
\rho(\xb)\ddot\ub(\xb,t)=\int_{{B_\delta(\mathbf{0})}}\left(\ut[\ub,\xb,t]\langle\xib\rangle+\ut[\ub,\xb+\xib,t]\langle -\xib\rangle\right) \ubM[\ub,\xb,t]\langle\xib\rangle\;d\xib +\bb(\xb,t), \\ 
 \text{ for }(\xb,t)\in\Omega\times[0,T],\label{eqn:peri_full}
\end{multline}
with boundary data supplied by
\begin{equation}
\ub(\xb,t)=\ub_{BC}(\xb,t) \text{, for }(\xb,t)\in\Omega_I\times[0,T].
\end{equation}
In \eqref{eqn:peri_full}, the identity $\ubM[\ub,\xb+\xib,t]\langle-\xib\rangle=-\ubM[\ub,\xb,t]\langle\xib\rangle$ has been used.
$\Omega_I:=\{\xb|\xb\in\real^d \backslash \omg,\,\text{dist}(\xb,\omg)<2\delta\}$ is the {\emph{interaction region}} in which boundary data $\ub_{BC}$ \YY{is prescribed}. \YY{Additionally, we denote $\tilde{\Omega}_I:=\{\xb|\xb\in\real^d \backslash \omg,\,\text{dist}(\xb,\omg)<\delta\}$.} With the peridynamic governing equation of motion, to obtain a unique solution $\ub$ for any forcing term $\bb$, nonlocal boundary conditions (``volume constraints'') must be prescribed on this interaction region.

The purpose of the present work is to learn the peridynamic material model, in the form of $\ut$, from training data in the form of loading/response function pairs $\{\ub^s\YY{(\xb,t)},\bb^s\YY{(\xb,t)}\}_{s=1}^S$. The material model involves neural operators and is not simply an algebraic expression. The learnt model \eqref{eqn:peri_full} is applied to solve for the displacement field $\ub(\xb,t)$ in new loading instances $\bb(\xb,t)$ distinct from the training instances. In addition to the displacement field, the peridynamic model provides other quantities of interest, such as the stress field given by:
\begin{equation}\label{eqn:peri_stress}
\bm{P}(\xb,t)=\int_{{B_\delta(\mathbf{0})}}\ubT[\ub,\xb,t]\langle\xib\rangle  \otimes \xib d\xib.
\end{equation}
Such an evaluation of the stress is especially important while learning from experimental measurements using DIC: while DIC has the major advantage to provide for full displacement fields, one of its main issues is that the stress field in heterogeneous state configurations cannot be directly accessed. By learning a constitutive law \eqref{eqn:peri_full} from DIC measurements, our method allows for estimating displacement, strain, and stress fields on different domain and loading instances.

\subsection{Nonlocal Neural Operators}

General nonlocal neural operators \citep{li2020neural,li2020multipole,li2020fourier,you2022nonlocal,you2022learning} were developed for scientific computing applications. These operators parameterize function-to-function mapping by incorporating a nonlocal operator \cite{du2013nonlocal} into neural network architectures. A prototypical instance seeks a PDE solution operator in material modeling problems, where the initial input field (body force/boundary load) is mapped to the corresponding displacement field via a nonlinear parameterized mapping. Using this solution operator concept, several architectures were developed in neural operator based methods \citep{li2020neural,li2020multipole,li2020fourier,you2022nonlocal,you2022physics,you2022learning,goswami2022physics,liu2023ino,gupta2021multiwavelet,lu2019deeponet,cao2021choose,yin2022continuous,hao2023gnot,li2022transformer,yin2022continuous}. 
Compared to classical neural networks that operate between finite-dimensional Euclidean spaces, one of the most remarkable advantages of neural operators is the capability to learn mappings between infinite-dimensional function spaces \citep{li2020neural,li2020multipole,li2020fourier,you2022nonlocal,Ong2022,gupta2021multiwavelet,lu2019deeponet,lu2021learning,goswami2022physics}. As a result, neural operators feature resolution independence, which implies that the prediction accuracy is invariant to the resolution of input functions. Furthermore, in contrast to classical PDE-based approaches, neural operators can be trained directly from data, and hence require no domain knowledge or pre-assumed PDEs. All these advantages make neural operators a promising tool for learning complex material responses from experimental measurements \citep{yin2022simulating,goswami2022physics,yin2022interfacing,you2022physics,li2020neural,li2020multipole,li2020fourier,lu2021comprehensive}.

Formally, a nonlocal neural operator aims to construct a surrogate operator $\mcG:\mbU\rightarrow \mbF$ that maps the input function $\ub(\xb)$ to the output function $\bb(\xb)$. 
The resolution-independence property is realized by parameterizing the layer update, $\mcJ$, as a nonlocal (integral) operator, given as:
\begin{equation}\label{eq:gkn}
\hb(\xb,l+1)=\mathcal{J}[\hb(\cdot,l)](\xb):=\sigma\left(\YY{\Wb}\hb(\xb,l)+\int_\omg \Kb(\xb,\qb;\vb)\hb(\qb,l) d\qb + \cb\right)\text{ .}
\end{equation}
Here, $\hb(\cdot,l)$ denotes the feature function of the $l$th layer, taking values in $\real^{d_h}$\YY{,} $\sigma$ is an activation function, $\YY{\Wb}\in\real^{d_h\times d_h}$, $\cb\in\real^{d_h}$ are trainable tensors parameterizing a point-wise linear transformation, and $\Kb\in\real^{d_h\times d_h}$ is a tensor kernel function whose parameters $\vb$ are to be learned. In earlier forms of nonlocal neural operators, the integral is extended to the whole set $\omg$. For improved efficiency, restrictions to a ball of radius $\delta$ centered at $\xb$, i.e. $B_\delta(\xb)$, can also be considered. However, as might be anticipated, this choice might compromise the accuracy of the method, since the support of Green's function generally spans the whole domain in PDE problems \cite{li2020neural}.

Despite the aforementioned advances of nonlocal solution operators, these approaches have limitations in physical consistency and domain/boundary condition generalizability. To learn the basic physical laws and generalize the solution, these operators require a large corpus of paired datasets, whose acquisition is very often \YY{prohibitive} in experiments. Moreover, the learnt operator is not applicable to different domain shapes/boundary conditions. To resolve these limitations, in our previous work \cite{jafarzadeh2024peridynamic}, the Peridynamic Neural Operator (PNO) was proposed, where we parameterize the nonlocal neural operator architecture as a nonlocal constitutive law based on the peridynamics formulation \eqref{eqn:peri_full}. As such, the model preserves the fundamental physical laws and is generalizable to different domain geometries and loading scenarios. In this context, the nonlocal neural operator aims to construct a surrogate operator $\mcG:\mbU\rightarrow \mbF$ that maps the displacement function $\ub(\xb)$ to the body load function $\bb(\xb)$:
\begin{equation}
  \mcG[\ub](\xb,t)\approx \rho(\xb)\ddot\ub(\xb,t)-\bb(\xb,t)\text{ ,}
  \label{eqn-Gapprox}
\end{equation}
where the operator $\mcG$ is formulated as:
\[\mcG[\ub](\xb,t):=\int_{{B_\delta(\mathbf{0})}}\left(\ut[\ub,\xb,t]\langle\xib\rangle+\ut[\ub,\xb+\xib,t]\langle -\xib\rangle\right) \ubM[\ub,\xb,t]\langle\xib\rangle\;d\xib.\]
Although $\bb$ was treated in Section~\ref{subsec-peri} as prescribed, here it becomes an output of the solution operator representing the external load that would be needed to make the approximation in \eqref{eqn-Gapprox} exact.
Herein, we parameterize the scalar force state $\ut$ with neural networks: \begin{equation}\label{eqn:umt}
  \ut[\ub,\xb,t]\langle{\xib}\rangle:= \sigma^{NN}(\omega(\xib), \vartheta(\xb,t), \ue[\ub,\xb,t]\langle\xib\rangle, |\xib|;\vb)\text{ ,}
\end{equation}
where
\begin{equation}\label{eqn:omega_aniso}
  \omega(\xib) :=\omega^{NN}(\xib;\wb)\text{ ,}
\end{equation}
\begin{equation}\label{eqn:dila}
  \vartheta(\xb,t) :=\dfrac{\int_{B_\delta (\mathbf{0})} \omega^{NN}\left(\xib;\wb\right)\ue[\ub,\xb,t]\langle\xib\rangle|\xib| d \xib}{\int_{B_\delta (\mathbf{0})} \omega^{NN}\left(\xib;\wb\right)|\xib|^2 d \xib}\text{ .}
\end{equation}
\YY{Here,} $\sigma^{NN}$ and $\omega^{NN}$ are scalar-valued functions that take the form of a (usually shallow) multi-layer perceptron (MLP) with learnable parameters $\vb$ and $\wb$, respectively. \YY{The symbol} $\omega$ denotes the kernel function characterizing the weighting of neighboring material points. $\vartheta$ is a nonlocal generalization of the dilatation, which describes the volume change of material near a point due to the deformation in volume. \YY{We emphasize that the global balance of linear momentum continues to hold regardless of whether the material model defining $\ubT$ is given by an algebraic expression or by a neural operator as in \eqref{eqn:umt}, since this dependence has no effect on the proof of \eqref{eqn-globalmomentum}.}

With the PNO architecture, one can model complex material mechanical responses learned from experimental data. In particular, given a set of observations $\mcD=\{\bb^s\YY{(\xb,t)},\ub^s\YY{(\xb,t)}\}_{s=1}^S$ of the loading field $\bb^s(\xb,t)$ and the corresponding displacement field $\ub^s(\xb,t)$, where $s$ is the sample index, the set of parameters in the network architecture is inferred by solving the following minimization problem
\begin{equation}\label{eqn:opt}
\min_{\vb,\wb}\mathbb{E}_{\ub}[C(\mcG[\ub;\vb,\wb],\mcG^\dag[\ub])]\approx \min_{\vb,\wb}\dfrac{1}{S}\sum_{s=1}^S[C(\mcG[\ub^s;\vb,\wb],\rho\ddot\ub^s-\bb^s)],
\end{equation}
where $C$ denotes a cost functional $C:\mbU\times\mbU\rightarrow\real$. Although $\ub^s\YY{(\xb,t)}$ and $\bb^s\YY{(\xb,t)}$ are (vector-valued) functions defined on a continuum, for numerical simulations, we assume that they are defined on a discretization of the domain defined as $\chi=\{\xb_1,\cdots,\xb_M\}\subset \omg$, and a temporal discretization as $t_n=n\Delta t$, $n=1,\cdots,N$. 
Once the constitutive law is obtained, for any new loading instance $\bb^{\text test}\YY{(\xb,t)}$, we solve for the displacement field $\ub\YY{(\xb,t)}$ using an iterative nonlinear static solver. The stress field $\bm{P}$ can also be calculated following \eqref{eqn:peri_stress}. \YY{In this work, we treat the horizon size, $\delta$, as a tunable hyperparameter which denotes the extent of nonlocal interaction. It should be noted that some expert-constructed nonlocal models have a corresponding local limit as $\delta\rightarrow 0$ \cite{tian2014asymptotically,you2020asymptotically,yu2021asymptotically}, and influence (weighting) functions can be defined as functions of $\delta$, allowing for different horizon sizes to reproduce the small-scale response, provided that the selected horizon are sufficiently small. 
However, as pointed out in \cite{du2020multiscale}, nonlocal formulations provide a more general class of models beyond local ones, and therefore they do not necessarily exhibit consistency for different values of $\delta$. In the case of our PNO, we aim to learn a nonlocal model with its full flexibility, and the learned model is neither required nor anticipated to give the same results when $\delta$ is changed.}




\section{Heterogeneous Peridynamic Neural Operators}\label{sec:heteroPNO}

\subsection{Mathematical Formulation}

\begin{figure}[!t]\centering
\includegraphics[width=.70\linewidth]{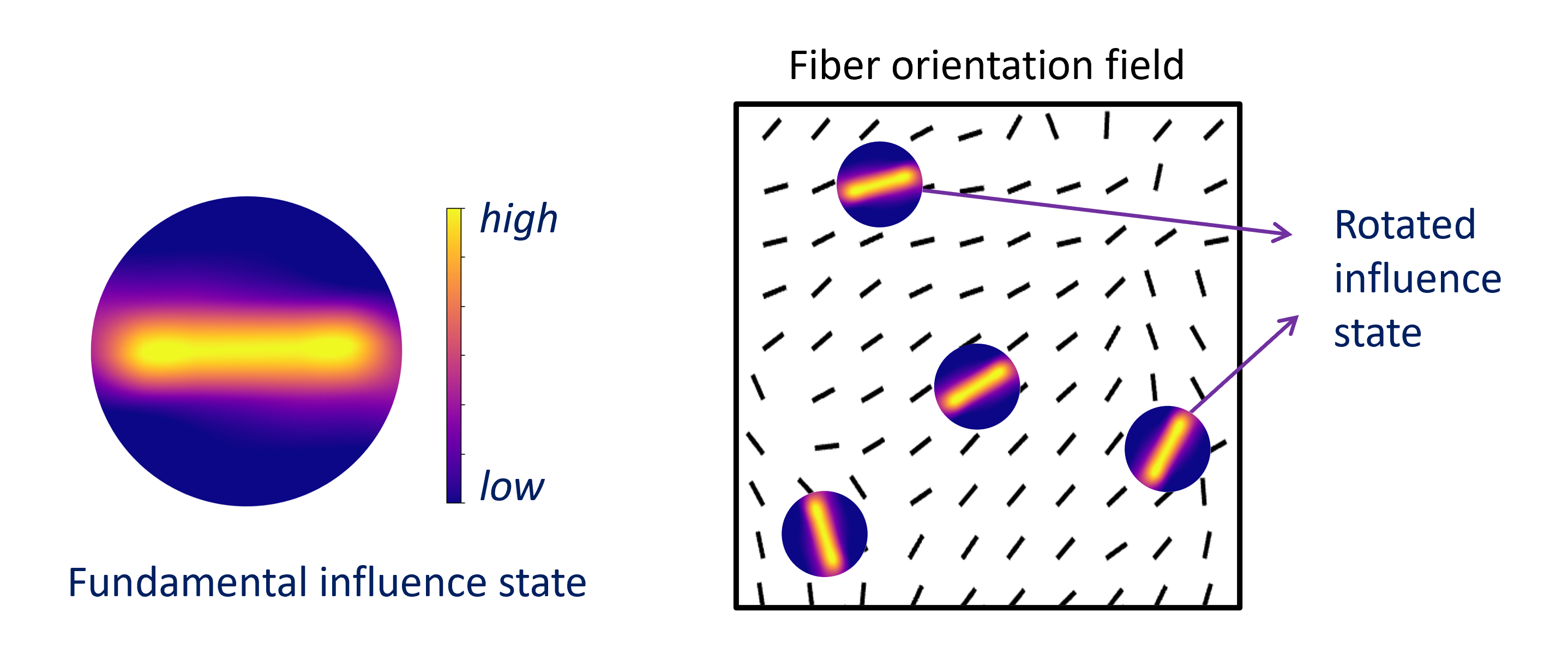}
 \caption{Schematic of the hetero-PNO mechanism, with a fundamental influence (kernel) function $\omega$ corresponding to a horizontal fiber orientation, which is rotated for each point independently, according to the local fiber angle.} 
 \label{fig:idea}
\end{figure}

In this section we introduce a heterogeneous PNO (HeteroPNO) architecture for the data-driven constitutive modeling of materials with position-dependent anisotropy. The particular materials of interest are biological tissues. These materials are soft and sustain large deformation. The position-dependent anisotropy in these tissues stems from embedded collagen fibers oriented in a heterogeneous pattern of directions in the plane. Figure \ref{fig:idea} (on the right) shows an example of such orientation field for a biological tissue.

The idea behind the new architecture is to assume that a fundamental influence (kernel) function $\omega(\xib)$ for an ideal homogeneous version of the tissue with unidirectional fiber orientation exists. This influence function can depend on direction of the bond vector $\xib$ as well as $|\xib|$. Then, in the case of the heterogeneous tissue, that influence state can be rotated at each location such that the stiffer direction of the influence state aligns with the fiber orientation in that location. Figure \ref{fig:idea} schematically illustrates this idea.

To construct such model, we replace \eqref{eqn:omega_aniso} with the following position-dependent influence function:
\begin{equation}\label{eqn:omega_heter}
  \omega(\xb, \xib) :=\omega^{NN}(\Rb(-\alpha(\xb)) \xib;\wb)\text{ ,}
\end{equation}
where
\begin{equation}\label{eqn:Rot_mat}
  \Rb(\theta) := 
\begin{bmatrix}
    \mathrm{cos} \theta & -\mathrm{sin} \theta \\
    \mathrm{sin} \theta & \mathrm{cos} \theta
\end{bmatrix}
\end{equation}
is the rotation matrix. The fundamental influence function has its stiffest direction along the $x$-axis ($\alpha = 0$). To account for the orientation of fibers in other directions at different positions,
each peridynamic bond $\xib$ is rotated clockwise by the fiber angle before evaluation. This is equivalent to rotating the fundamental influence function through an angle $\alpha$ counterclockwise at each point to align with the actual fiber orientation on that point. 
\YY{Although one can alternatively choose the more general form: $\omega(\xb, \xib) :=\omega^{NN}(\alpha(\xb), \xib;\wb)$, the form in \eqref{eqn:omega_heter} enjoys the encoded knowledge of periodicity with respect to $\alpha$, i.e., $\omega(R(\alpha)\xib)=\omega(R(\alpha+2\pi)\xib)$.} 

The position-dependent influence state is then used in the HeteroPNO architecture as follows:
\begin{equation}\label{eqn:balance}
\mcG[\ub](\xb,t)=\int_{{B_\delta(\mathbf{0})}}\left(\ut[\ub,\xb,t]\langle\xib\rangle+\ut[\ub,\xb+\xib,t]\langle -\xib\rangle\right) \ubM[\ub,\xb,t]\langle\xib\rangle\;d\xib=\rho(\xb)\ddot{\ub}(\xb,t)-\bb(\xb,t),
\end{equation}
where
\begin{align}
  \ut[\ub,\xb,t]\langle{\xib}\rangle:=&\sigma^{NN}(\omega(\xb, \xib), \vartheta(\xb,t), \ue[\ub,\xb,t]\langle\xib\rangle, |\xib|;\vb)\;\text{ ,}\label{eqn:umt_het}\\
  \omega(\xb, \xib) :=&\omega^{NN}(\Rb(-\alpha(\xb)) \xib;\wb)\text{ ,}\label{eqn:omega}\\
  \vartheta(\xb,t) :=&\dfrac{\int_{B_\delta (\mathbf{0})} \omega^{NN}\left(\xb, \xib;\wb\right)\ue[\ub,\xb,t]\langle\xib\rangle|\xib| d \xib}{\int_{B_\delta (\mathbf{0})} \omega^{NN}\left(\xb, \xib;\wb\right)|\xib|^2 d \xib}\text{ .}\label{eqn:dila_het}
\end{align}
Our goal is to learn the optimal influence (kernel) function $\omega^{NN}$, the force state $\sigma^{NN}$, and the pointwise fiber orientation function $\alpha(\xb)$ from data.

\subsection{Machine learning algorithm}\label{sec:algorithm}
In this section we describe how to train the HeteroPNO models for two different scenarios: 1) learning the constitutive model from displacement field--force field data pairs, with the ground-truth fiber orientation field given. 2) Discovering both the constitutive law \textit{and} the microstructure (fiber orientation field) from displacement-force field data.

\begin{algorithm}
\caption{Algorithm for HeteroPNOs.}\label{alg:ML}
\begin{algorithmic}[1]
\State \textbf{Pre-processing Phase:}
\State Read data and inputs: $\chi=\{\xb_i\}_{i=1}^M$,
\If{Case I}
    \State $\mcD_{tr}:=\{\ub^{s,tr}(\xb_i),\bb^{s,tr}(\xb_i)\}_{i=1,s=1}^{M,S_{tr}}$ for training, $\mcD_{val}:=\{\ub^{s,val}(\xb_i),\bb^{s,val}(\xb_i)\}_{i=1,s=1}^{M,S_{val}}$ for validation sets.
\EndIf
\If{Case II}
    \State $\mcD_{tr}:=\{\ub^{s,tr}(\xb_i)\}_{i=1,s=1}^{M,S_{tr}}, \{\bm{P}^{s,tr}\}_{s=1}^{S_{tr}}$ for training, $\mcD_{val}:=\{\ub^{s,val}(\xb_i)\}_{i=1,s=1}^{M,S_{val}}, \{\bm{P}^{s,val}\}_{s=1}^{S_{val}}$ for validation sets.
\EndIf
 
\State Find and store the peridynamic neighbor node set $N(\xb_i)$ in the form of message passing graph structure:
\For{$i = 1:M$}
    \State Find all $\xb_j\in\chi$ such that $|\xb_j - \xb_i|< \delta$.
\EndFor
\State \textbf{Training Phase:}
\For{$ep = 1: epoch_{max}$}
    \State Initialize error: $E_{tr} = 0$
    \For{ each batch}
        \State Reset batch loss: $loss = 0$
        \If{Case I}
            \State Compute batch loss following \eqref{eqn: caseI_loss}.
        \ElsIf{Case II}
            \State For each training sample in this batch, run nonlinear solver to obtain $\mcG^{-1}[\bb^s]$.
            \State Compute peridynamic stress for all training samples from \eqref{eqn:peri_stress} and get the averaged axial components $\hat{\bm{P}}={\hat{P_{11}}, \hat{P_{22}}}$.
            \State Compute batch loss following \eqref{eqn: caseII_loss}.
        \EndIf

        \State Update HeteroPNO parameters $\wb$, $\vb$, and $\bm{\alpha}$ with the Adam optimizer.
    \EndFor
    \State Compute training error $E_{tr}$.
    \State \textbf{Validation Phase:}
    \If{training error decreases}
        \State Initialize validation error: $E_{val} = 0$
        \If{Case I}
            \State Compute validation error following \eqref{eqn: caseI_loss}.
        \ElsIf{Case II}
            \State For each validation sample, run nonlinear solver to obtain $\mcG^{-1}[\bb^s]$.
            \State Compute peridynamic stress from \eqref{eqn:peri_stress} and get the averaged axial components $\hat{\bm{P}}={\hat{P_{11}}, \hat{P_{22}}}$.
            \State Compute validation error following \eqref{eqn: caseII_loss}.
        \EndIf

        \State If validation error also decreases, save the current model. 
    \EndIf
\EndFor
\end{algorithmic}
\end{algorithm}

\paragraph{\textbf{Constitutive model learning.}} 

In this work we focus on (quasi)static, two-dimensional ($d=2$) applications, although the prescribed architecture is readily applicable to higher-dimensional domains and dynamic settings. We aim to learn the surrogate operator $\mcG$ such that
$$\mcG[\ub](\xb)\approx -\bb(\xb)\text{ ,}$$
where $\bb$ is prescribed.
Algorithm \ref{alg:ML} provides the structure of the pseudo code used for this work to learn the constitutive model (and microstructure if applicable). The program is coded in Python and uses the PyTorch libraries for GPU computing and the Adam optimizer for minimizing the loss function. In all tests, the neural network functions for $\omega^{NN}$ and $\sigma^{NN}$ are multi-layer perceptrons (MLP) with three layers and ReLu as the activation function. The widths of the MLP layers are denoted by $(W_{in}, W_{h1}, W_{h2}, W_{out})$, where $W_{in}$ is the input's dimension, $W_{h1}, W_{h2}$ are the widths of the hidden layers, and $W_{out}$ is the dimension of the output. For all of the present examples, $\omega^{NN}$, $W_{in}=2$ and $W_{out}=1$. However, $W_{out}$ can take on larger values in other applications. From \eqref{eqn:umt_het}, since $|\xib|$ and $\ue$ are scalars and $\vartheta(\xb)$ has the same dimension as $\omega(\xib)$ (see \eqref{eqn:dila}), it follows that $W_{in}=4$ and $W_{out}=1$ in the $\sigma^{NN}$ MLP. 
$W_{h1}, W_{h2}$ in both $\omega^{NN}$ and $\sigma^{NN}$ as well as the horizon size $\delta$ are hyperparameters of choice to tune.

Formally, we assume that measurements are available in the following format:
\begin{itemize}
    \item Coordinates of nodes where measurements are taken: $\chi=\{\xb_i\}_{i=1}^M$, where $M$ is the total number of nodes.
    \item External force-displacement function pairs at each node $\xb_i$ and for each sample $s$:  
    $\{\ub^{s}_i,\bb^s_i\}_{i=1,s=1}^{M,S}$ where $S$ is the total number of samples.
    \item If the external forces are all zero (which is the case for the biaxial tension experiments described below), we can compare the area-averaged stress components given by \eqref{eqn:peri_stress} against the measured loads in the testing machine: $\{\bm{P}^s\}_{s=1}^{S}$ where $\bm{P}^s = \{P^s_{11}, P^s_{22}\}$ and $S$ is the total number of samples.
\end{itemize}
For the purposes of training, validation and test, we split the available data into three sets: $S_{tr}$, $S_{val}$, $S_{test}$ denote the total number of samples in training, validation, and test sets, respectively. 
Depending on whether the data includes nonzero external forces (Case I) or not (Case II), 
either of the following two different loss functions is used. 

In Case I, we follow the convention of the nonlocal neural operator literature \cite{li2020neural,li2020fourier} and consider the relative $L^2$ error of the output function, $\bb$, as the loss function, which is given by
\begin{equation}\label{eqn: caseI_loss}
    \text{loss}_b = \frac{1}{S_{tr}}\sum_{s = 1}^{S_{tr}}\dfrac{\vertii{\mcG[\ub^s]+\bb^s}_{l^2(\omg)}}{\vertii{\bb^s}_{l^2(\omg)}}\text{ .} 
\end{equation}
Physically, the loss function in \eqref{eqn: caseI_loss} measures the error in the internal force distribution on the nodes needed to equilibrate the system.

If external forces are zero (Case II), the loss function \eqref{eqn: caseI_loss} fails since the denominator becomes zero. \YY{Given zero external forces, removing the denominator would lead to a loss function of form: $\frac{1}{S_{tr}}\sum_{s = 1}^{S_{tr}}{\vertii{\mcG[\ub^s]}_{l^2(\omg)}}$. Then, a neural network with $\sigma^{NN}=0$ yields a zero loss. That means, removing the denominator would result in a trivial solution. Such a model would return zero forces for any deformation and the constitutive model will not be learnt.} To circumvent this problem, in Case II, we define the loss in terms of the displacement field and averaged axial first Piola-Kirchhoff stress components:
\begin{equation}\label{eqn: caseII_loss}
   loss_u = \beta\left(\frac{1}{S_{tr}}\sum_{s = 1}^{S_{tr}}\dfrac{\vertii{\mcG^{-1}[\bb^s]-\ub^s}_{l^2(\omg)}}{\vertii{\ub^s}_{l^2(\omg)}}\right) + (1 - \beta)\left(\frac{1}{S_{tr}}\sum_{s = 1}^{S_{tr}}\dfrac{\verti{\hat{\bm{P}^s}-\bm{P}^s}}{\Bar{\bm{P}}}\right)\text{ ,}
\end{equation}
where $\hat{\bm{P}^s}$ denotes the spatial average of axial 1st Piola-Kirchhoff stresses for sample $s$, and $\Bar{\bm{P}}$ is the mean of axial ground truth stresses across all training samples. This results in a scaled absolute error measure for stress contribution to the loss. \YY{In \eqref{eqn: caseII_loss}, }$\mcG^{-1}[\bb^s]$ denotes the numerical solution of $\mcG[\ub]=-\bb^s$ using an iterative nonlinear static solver (see line 18 in Algorithm~\ref{alg:ML}). In this work we use the Polak-Ribiere conjugate gradient method \cite{shewchuk1994introduction}, due to its efficiency and robustness in peridynamics nonlinear problems \cite{van2014relationship}. 
Additional inputs needed for the solver are ${\ub^{s}_0}$, ${\ub^{s}_{BC}}$, $tol$, $itr_{max}$, which denote the initialization displacement (which may or may not be zero), Dirichlet boundary condition values on $\Omega_I$, convergence tolerance, and maximum allowed iterations, respectively. \YY{In \eqref{eqn: caseII_loss}, }$\beta$ takes values between 0 and 1 and is a hyperparameter to tune. It determines how much the learning relies on stress versus the displacement predictions.

As discussed in \cite{jafarzadeh2024peridynamic}, Case II is computationally more expensive and has significantly higher memory requirements. \YY{In Case II, as observed from Algorithm~\ref{alg:ML}, line 18, the nonlinear solver is called for each data instance of each epoch. As the most straightforward and systematic approach to differentiate the loss function for learning, we have used automatic differentiation (AD). This approach, however, needs the storage of intermediate steps in the iterative solver that occur within each optimizer step, leading to a higher memory requirement for Case II.} To make the Case II algorithm affordable, the following strategies are taken to reduce the number of solver's iterations:
\begin{enumerate}
    \item Avoiding the use of too small tolerances for convergence criterion.
    \item Smart initialization of the solution. Here, in particular, we used the displacements from another data sample with a close temporal sequence as the initial guess for the solver. 
    \item Limiting the maximum number of iterations to the lowest possible value at which convergence for most samples is likely to be achieved.
\end{enumerate}
As described in \cite{jafarzadeh2024peridynamic}, in Case II, we found it helps to modify the architecture of the scalar force state given by \eqref{eqn:umt} as follows:
\begin{equation}
    \ut[\ub,\xb]\langle{\xib}\rangle:= \sigma^{NN}(\omega(\xb, \xib), \vartheta(\xb), \ue[\ub,\xb]\langle{\xib}\rangle, |\xib|;v) \; e(\xib, \etab)\text{ .} 
    \label{eqn:umtcaseii}
\end{equation}
which guarantees that a small deformation bond will be associated with a small force state, and it causes the larger deformation bonds to weigh more on the overall equilibrium deformation. Thus, the modified form \eqref{eqn:umtcaseii} tends to improve the training efficiency. \YY{To avoid PNO models that are sensitive to initialization and may have multiple solutions depending on initialization, the smart initialization strategy is employed only in the training samples. For the validation and test samples we instead solve for the displacement field with a zero initialization.  This validation strategy selects the best model based on robust convergence from a standard/not-smart initialization (zero) on the validation set. This process is anticipated to avoid models with multiple solutions and those that are non-invertible.}

\bigskip
\paragraph{\textbf{Microstructure Discovery.}} 

When the microstructure field in the form of $\alpha(\xb)$ is not provided, we \YY{further parameterize it as a shallow neural network $\alpha^{NN}(\xb)$} to be found together with other HeteroPNO's learnable MLPs' weights and biases. Thus, we discover the $\alpha$ field from displacement-force/stress data only. A two-stage training strategy is developed. First, we train a homogeneous PNO (HomoPNO). We then take the trained HomoPNO force state function ($\sigma^{NN}$) as the initial $\sigma^{NN}$ for the HeteroPNO, and train for $\sigma^{NN}$, $\omega^{NN}$, and $\alpha^{NN}(\xb)$ simultaneously. For the purpose of verification, in our numerical experiments we consider two scenarios: 
\begin{itemize}
\item Using actual fiber angles, $\alpha(\xb)$, in \eqref{eqn:omega_heter} as known information. This scenario will be denoted as HeteroPNO I. 
\item Treating $\alpha(\xb)$ as an unknown to be learned from data. This scenario will be denoted as HeteroPNO II.
\end{itemize}

\bigskip
\paragraph{\textbf{Stress field calibration.}} 
{\YY{
Unlike Case II, where stress information is directly incorporated in the loss function, the Case I loss function only enforces $\mcG[\ub](\xb)= -\bb(\xb)$. 
This results in a nonuniqueness in the determination of the scalar force state in Case I, since the loss function
does not incorporate data on the stress.
To address this, we modify the HeteroPNO to account for the average stress tensor within the sample,
which is easily determined from the measured total loads on the edges.
Suppose the HeteroPNO force state, $\ut[\xb]$ is given.
Let $\bm{P}$ be the mean stress in the sample, as determined by the total loads.
We seek to enforce the condition
\[  \frac{1}{A}\int_\Omega\int_{B_\delta(0)}(\tilde\ut[\ub,\xb,t]\,\ubM[\ub,\xb,t])\langle\xib\rangle \otimes \xib\;d\xib\;d\xb=\bm{P}  \]
where $A$ is the area of the speciment and $\tilde\ut$ is the scalar force state after correction.
Such a correction is provided by assuming there is are constants $C_1$ and $C_2$, independent of $\xb$, such that
\[   \tilde\ut\langle\xib\rangle = \ut\langle\xib\rangle +C_1M_1^2+C_2M_2^2 \]
where $M_1$ and $M_2$ are the components of $\ubM\langle\xib\rangle$.
From the geometry of the experiment, we know that $P_{12}=P_{21}=0$.
Then, using \eqref{eqn:peri_stress},
\begin{eqnarray*}
    P_{11}&=&\frac{1}{A}\int_\Omega\int_{B_\delta(0)}\ut\langle\xib\rangle M_1^2|\xib|\;d\xib\;d\xb + 
    \int_{B_\delta(0)}(C_1M_1^2+C_2M_2^2)M_1^2|\xib|\;d\xib, \\
    P_{22}&=&\frac{1}{A}\int_\Omega\int_{B_\delta(0)}\ut\langle\xib\rangle M_2^2|\xib|\;d\xib\;d\xb + 
    \int_{B_\delta(0)}(C_1M_1^2+C_2M_2^2)M_2^2|\xib|\;d\xib,
\end{eqnarray*}
which form a linear algebraic system in which the only unknowns are $C_1$ and $C_2$. As such, we obtained the calibrated force state $\tilde\ut[\ub,\xb,t]$, which will be employed to calculate the stress field in Case I.
}
}

\section{Verification on Synthetic Dataset}\label{sec:HGO}

\begin{figure}[!t]\centering
\includegraphics[width=.99\linewidth]{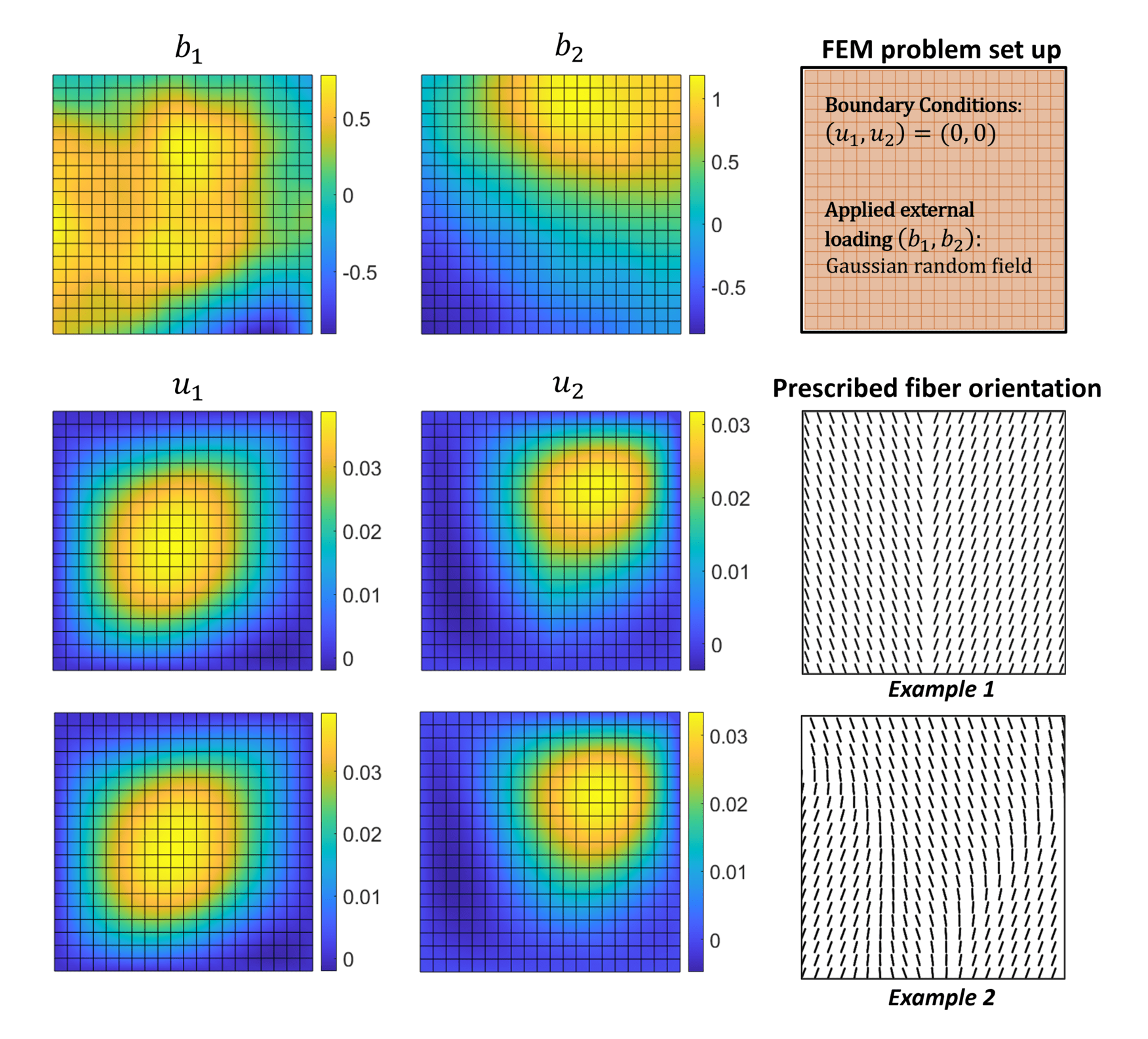}
 \caption{\YY{Demonstration of the heterogeneous synthetic data generation, under HGO constitutive law and heterogeneous fiber orientation field. Top row: FEM boundary value problem set up (right) and one instance of the random applied external forces (left). Middle and bottom rows: prescribed collagen fiber orientation in the two examples (right), and the resulting displacement fields (left) associated with the external forces shown on the top.}} 
 \label{fig:HGO_data_gen}
\end{figure}

In this section, we illustrate the performance of the proposed HeteroPNO on a benchmark material modeling problem. In particular, we consider a synthetic dataset describing the deformation of a hyperelastic and anisotropic fiber-reinforced material. All our numerical experiments were performed on a machine with a single Nvidia RTX 3090 GPU. 

\subsection{Data preparation}

To generate the training and test samples, the Holzapfel-Gasser-Odgen (HGO) model \cite{holzapfel2000new} was employed to describe the constitutive behavior of the material in this example, with its strain energy density function given as:
\begin{align*}
    \psi & = \frac{E}{4(1+\nu)}(\overline{I}_{1} - 2) - \frac{E}{2(1+\nu)}\ln(J)\\ 
    & +\frac{k_{1}}{2k_{2}}\left(\exp{(k_{2}\langle S(\alpha) \rangle^{2}}) + \exp{(k_{2}\langle S(-\alpha) \rangle^{2}}) - 2\right) + 
    \frac{E}{6(1-2\nu)}\left( \frac{J^{2} - 1}{2} - \ln{J} \right).
\end{align*}
Here, $\langle \cdot \rangle$ denotes the Macaulay bracket, and the fiber strain of the two fiber groups is defined as:
\begin{equation*}\label{eqn:fiberstrain}
    S(\alpha) = \frac{\overline{I}_{4}(\alpha) - 1 + |\overline{I}_{4}(\alpha) - 1|}{2}.
\end{equation*}
where $k_{1}$ and $k_{2}$ are fiber modulus and the exponential coefficient, respectively, $E$ is the Young's modulus for the non-fibrous ground matrix, and $\nu$ is the Poisson's ratio. Moreover, $\overline{I}_{1}=\text{tr}(\Cb)$ is the first invariant of the right Cauchy-Green tensor $\Cb=\Fb^T\Fb$, $\Fb:=\Fb(\ub)=\Ib+\dfrac{\partial \ub}{\partial \xb}$ is the deformation gradient tensor, and $J = \det \Fb$. For the fiber group with angle direction $\alpha$ from the reference direction, $\overline{I}_{4}(\alpha)=\mathbf{n}^T(\alpha)\Cb\mathbf{n}(\alpha)$ is the fourth invariant of the right Cauchy-Green tensor $\Cb$, where $\mathbf{n}(\alpha)=[\cos(\alpha), \sin(\alpha)]^{T}$. Here, $\alpha$ denotes the collagen fiber orientation angle.

In this problem, the specimen is assumed to be subject to different body loads $\bb(\xb)$, and the goal is to find the corresponding displacement field $\ub:\Omega\rightarrow\real^2$ under each body loading, where $\Omega:=[0,1]^2$. To generate the high-fidelity (ground-truth) dataset, we sampled $250$ different body loads $\bb(\xb)$ from a random field, following the algorithm in \cite{LangPotthoff2011,yin2022interfacing}. \YY{To include heterogeneity in the generated data, we consider two exemplar fiber orientation patterns. In the first example, we assume that the left and the right halves of the square domain have fibers along 110$^{\text{o}}$ and 70$^{\text{o}}$ directions, respectively, with a sharp transition along the central vertical line (see the second row of Figure \ref{fig:HGO_data_gen}). In the second example, we consider a smooth fiber orientation pattern with less symmetry and more complexity generated as a Gaussian random field (see the second row of Figure \ref{fig:HGO_data_gen}). In both examples, the external force, $\bb(\xb)$ is taken as the restriction of a 2D random field, $\phi(\xb) = \mathcal{F}^{-1}(\gamma^{1/2}\mathcal{F}(\Gamma))(\xb)$. Here, $\Gamma(\xb)$ is a Gaussian white noise random field on $\real^2$, $\gamma=(w_1^2+w^2_2)^{-\frac{5}{4}}$ represents a correlation function, $w_1$ and $w_2$ are the wave numbers on $x$ and $y$ directions, respectively, and $\mathcal{F}$ and $\mathcal{F}^{-1}$ denote the Fourier transform and its inverse, respectively. As such, this random field is anticipated to have a zero mean and covariance $C(\xb,\yb)=\int \exp(-2\pi i \bm{\omega}\cdot(\xb-\yb))\gamma(\bm{\omega})d\bm{\omega}$. For the detailed implementation of Gaussian random field sample generation, we refer interested readers to \cite{lang2011fast}.} Then, for each sampled traction loading, we solved the displacement field on the entire domain by solving the weak form of displacement formulation for the equilibrium: find the displacement field $\ub(\xb)\in \Ub_0:=\{\wb(\xb)\in[H^1([0,1]^2)^2|\wb(\xb)=0 \text{ on }\partial \Omega]\}$ which satisfies:
\begin{align*}
\int_\Omega \sigmab:\dfrac{1}{2}\left(\left(\dfrac{\partial \vb}{\partial \xb}\right)^T\Fb(\ub)+\Fb(\ub)^T\left(\dfrac{\partial \vb}{\partial \xb}\right)\right) d\xb-\int_\Omega\rho\bb\cdot\vb d\xb=0,\vb\in\Ub_0,
\end{align*}
where $\sigmab:=\dfrac{\partial \psi}{\partial \Eb}$ is the second Piola-Kirchhoff stress tensor, and $\Eb:=\dfrac{1}{2}\left(\left(\dfrac{\partial \ub}{\partial \xb}\right)^T+\left(\dfrac{\partial \ub}{\partial \xb}\right)+\left(\dfrac{\partial \ub}{\partial \xb}\right)^T\left(\dfrac{\partial \ub}{\partial \xb}\right)\right)$ is the Green-Lagrange strain tensor. \YY{Note that $\xb$ in this formulation denotes the position vector at the reference (initial) configuration.} In particular, we use the finite element method implemented in FEniCS \citep{alnaes2015fenics}, with the displacement field approximated by continuous piecewise linear finite elements with rectangular mesh, and the grid size taken as $0.05$. \YY{This grid size corresponds to an error of less than 2\% relative to the reference solution obtained from a super-resolution simulation with a 100,000 nodes. We used the default Newton method's parameters in FEniCS for solving the nonlinear variational problem: a maximum iteration of 50, a relative tolerance of 1e-9, and an absolute tolerance of 1e-10}. Once the finite element solution was obtained, it was interpolated onto $\chi$, a structured $21 \times 21$ grid which will be employed as the discretization in our HeteroPNO. 

Figure \ref{fig:HGO_data_gen} shows the problem setting and the force and the displacement fields for one of the generated samples. \YY{Note that PNO is a nonlocal formulation and therefore generally requires prescribed displacement boundary data $\ub_{BC}$ in a nonlocal boundary region $\Omega_I$ of size $2\delta$. Denoting the data domain, where the displacement measurements are provided, as $\Omega_{data}$, with the purpose of understanding the boundary effects we consider two approaches. In the first approach, we set the learning domain as $\Omega=\Omega_{data}$, then extrapolate the displacement field to obtain the datum in $\Omega_I$ following \cite{jafarzadeh2024peridynamic}. In particular, we extend the domain on all four edges for each sample by $2\delta$ using the mirror-based fictitious nodes method \cite{zhao2020algorithm} for peridynamic computations. Note that the mirror-based fictitious nodes method \cite{zhao2020algorithm}, as well as other local-to-nonlocal boundary condition methods \cite{d2022prescription,yu2021asymptotically}, are generally developed for homogeneous materials. Hence, they introduce unavoidable errors in the displacement boundary data $\ub_{BC}$, which would potentially introduce discrepancies in the learnt fiber orientation field. This approach will be denoted as ``FNM'' (fictitious nodes method). In the second approach, we assume that measurements are also given on  the boundary region $\Omega_I$, and then $\Omega_{data}=\omg\cup\omg_I$. We then impose the true volume constraint on $\Omega_I$. As such, we 
avoid the possible additional errors but it would potentially waste some measurements on prescribing boundary. Therefore, it is less preferred in data scarce problems, such as the present bio-tissue application, in which the DIC measurements are provided on a sparse grid. In the following discussion, this approach will be denoted as ``VC'' (volume constraint).}

\subsection{Learning the constitutive law and microstructure}

\begin{table}[]
\begin{tabular}{|c|c|}
\hline
Hyperparameter & Values \\
\hline
Learning rate & {0.03, 0.01, 0.003, 0.001} \\
\hline
Decay rate & {0.9, 0.7, 0.5} \\
\hline
Weight decay & {0, 1e-5, 1e-4} \\
\hline
Maximum epoch & {500} \\
\hline
Batch size & {5} \\
\hline
\end{tabular}
\caption{\YY{List of hyperparameters used with Adam optimizer to tune the PNO models for synthetic dataset, examples 1 and 2.}}
\label{table:HGO_hyperparam}
\end{table}

\begin{table}[]
\begin{tabular}{|c|c|c|c|c|c|c|}
\hline
Example&\multicolumn{3}{c|}{HGO (Case I)}&\multicolumn{3}{c|}{Tissue (Case II)}\\
\hline
Setting & HomoPNO  & HeteroPNO I  & HeteroPNO II  & HomoPNO  & HeteroPNO I  & HeteroPNO II \\
\hline
 $\#$ of parameters & 398.6 k & 398.6 k &  415.6 k & 6.82 k & 6.82 k & 23.8 k\\
\hline
Memory (MB) & 23.06 & 23.06 & 23.34 &  2869.1 & 2922.6 & 3247.6 \\
\hline
 Time per epoch (s) & 2.23 & 2.91 & 3.04 & 64.6 & 64.8 & 60.2 \\
\hline
\end{tabular}
\caption{\YY{Number of trainable parameters, training time per epoch, and allocated GPU memory for the PNO models under the two different settings: Case I (synthetic dataset examples), Case II (biotissue dataset example).}}
\label{table:time_memory}
\end{table}

\begin{table}[]
\begin{tabular}{|c|c|c|c|c|}
\hline
\multicolumn{2}{|c|}{error} & training (200) & validation (25) & test(25) \\
\hline
\multirow{3}{*}{force} & HomoPNO & 16.02\% & 17.05\% & 16.33\% \\
 & HeteroPNO I (sharp $\alpha$)& {8.98}\% & {9.56}\% &{8.75}\% \\
  & HeteroPNO I (smoothed $\alpha$)& {6.27}\% & {6.65}\% &{6.18}\% \\
  & HeteroPNO II (learned $\alpha$)& {6.46}\% & {6.92}\% &{6.27}\% \\
\hline
\multirow{3}{*}{displacement} & HomoPNO & 6.45\% & 6.80\% & 6.91\% \\
 & HeteroPNO I (sharp $\alpha$)& {1.77}\% & {1.91}\% &{1.81}\% \\
  & HeteroPNO I (smoothed $\alpha$)& {1.21}\% & {1.21}\% &{1.41}\% \\
  & HeteroPNO II (learned $\alpha$)& {1.27}\% & {1.32}\% &{1.24}\% \\
\hline
\end{tabular}
\caption{\YY{Synthetic dataset, example 1: Averaged relative $L^2$-norm error for HomoPNO and HeteroPNO predictions on forces (given displacement) and on displacement (given boundary conditions).}}
\label{table:HGO_errors}
\end{table}

\begin{figure}[!t]\centering
\includegraphics[width=.9\linewidth]{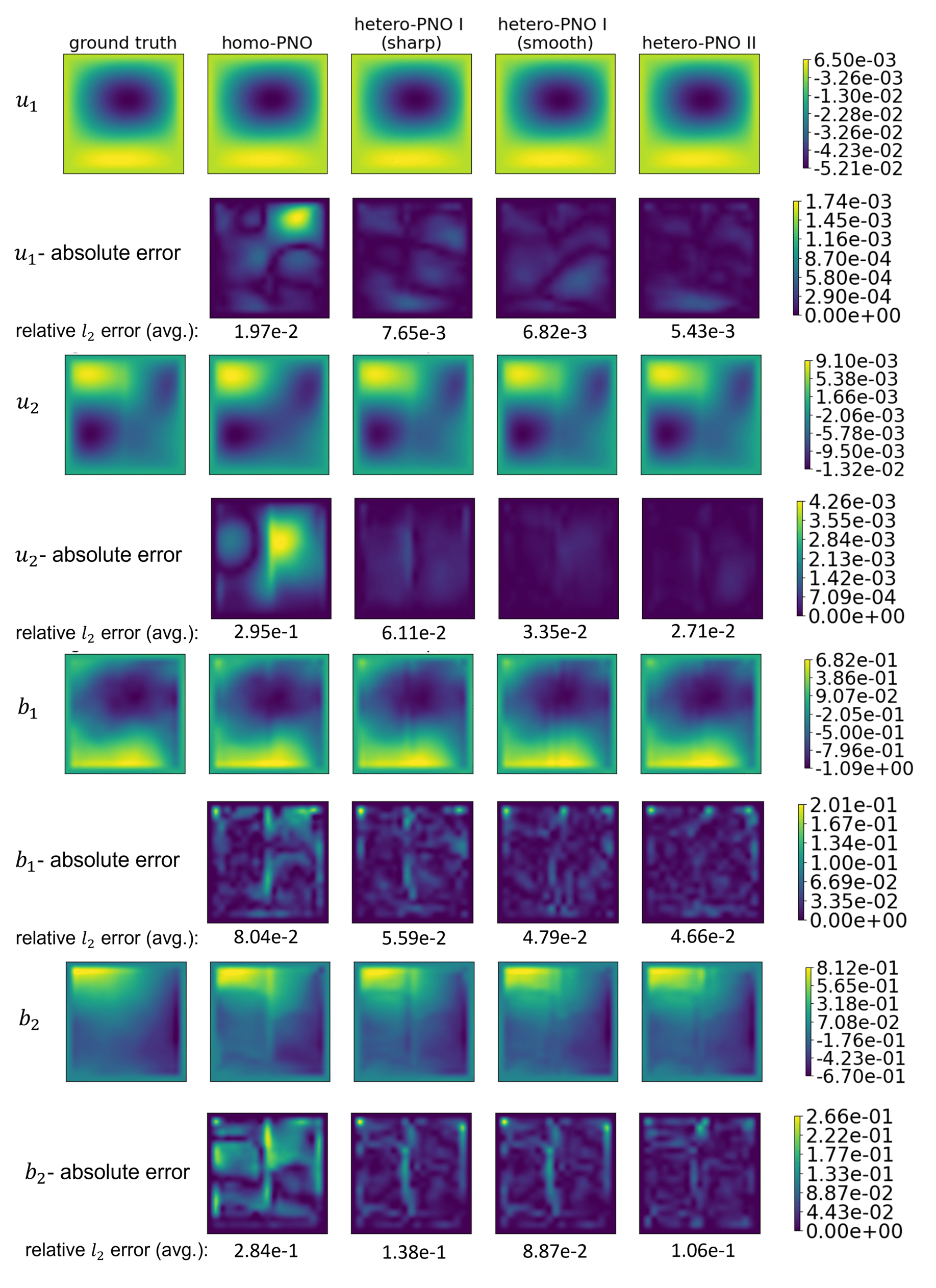}
 \caption{\YY{Synthetic dataset, example 1: HomoPNO and HeteroPNO predictions of displacement field (given external load and boundary conditions), and external forces (given displacement field) against the ground truth.}}
 \label{fig:HGO_UB_results}
\end{figure}

\begin{figure}[!t]\centering
\includegraphics[width=.9\linewidth]{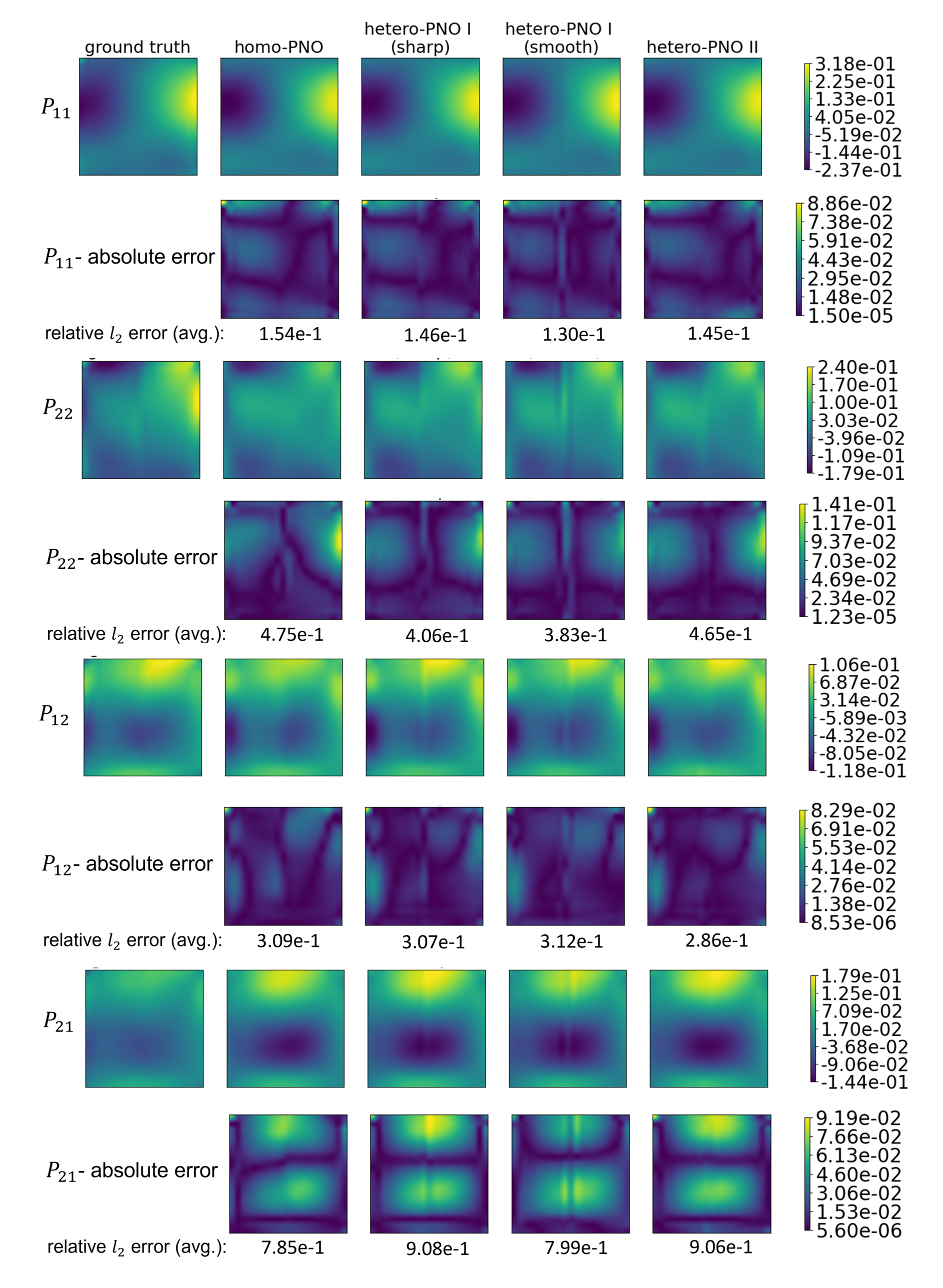}
 \caption{\YY{Synthetic dataset, example 1: HomoPNO and HeteroPNO predictions of the first Piola-Kirchhoff stress tensor against the ground truth.}}
 \label{fig:HGO_stress}
\end{figure}

\begin{figure}[!t]\centering
\includegraphics[width=.89\linewidth]{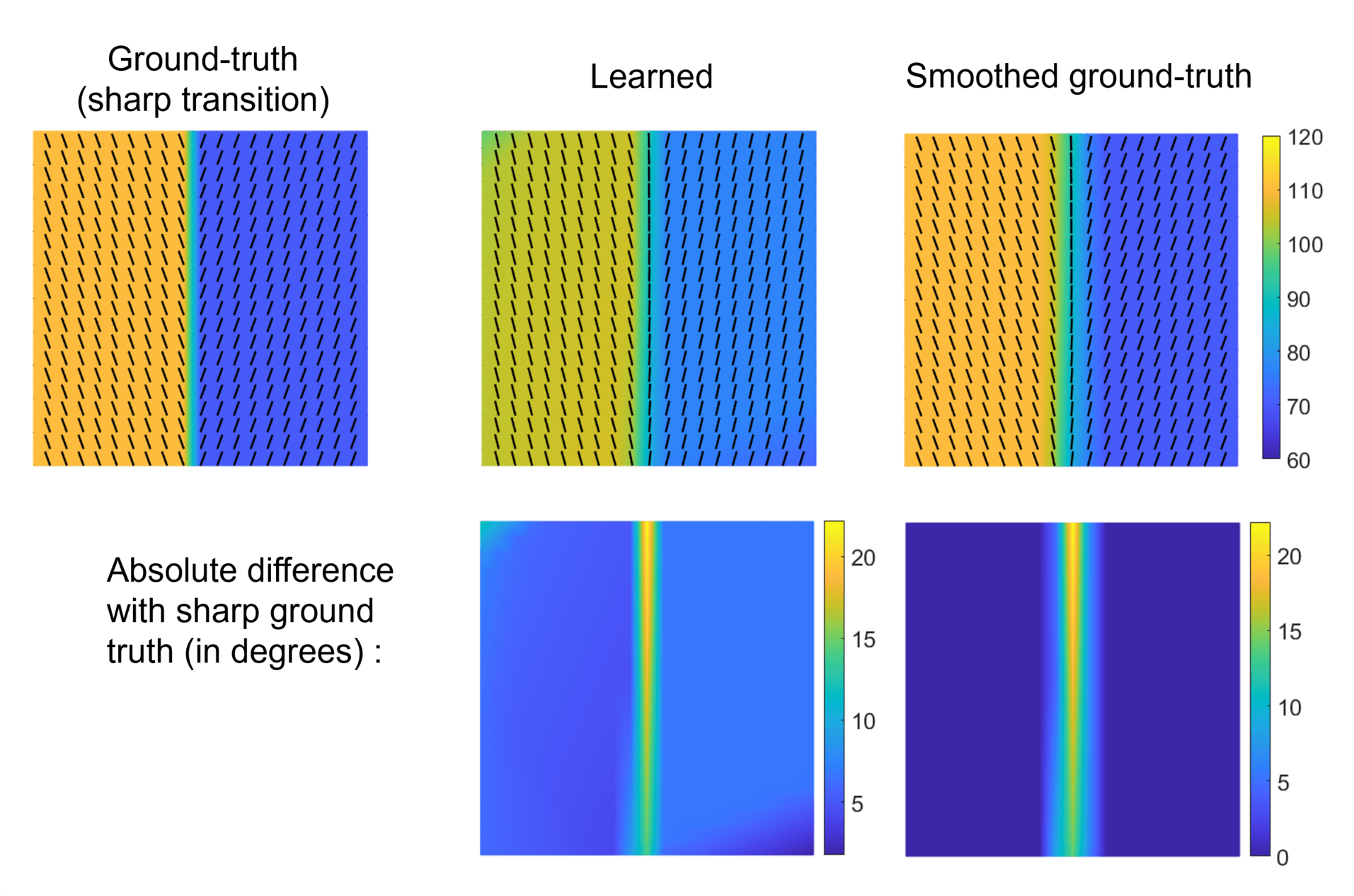}
 \caption{\YY{Synthetic dataset, example 1: Discovered hidden fiber orientation by Hetero-PNO on the synthetic data set and comparison against the ground truth. The averaged absolute error for the learned orientation map is \YY{6.55$^{\text{o}}$}.}}
 \label{fig:HGO_fiber}
\end{figure}

\begin{figure}[!t]\centering
\includegraphics[width=.79\linewidth]{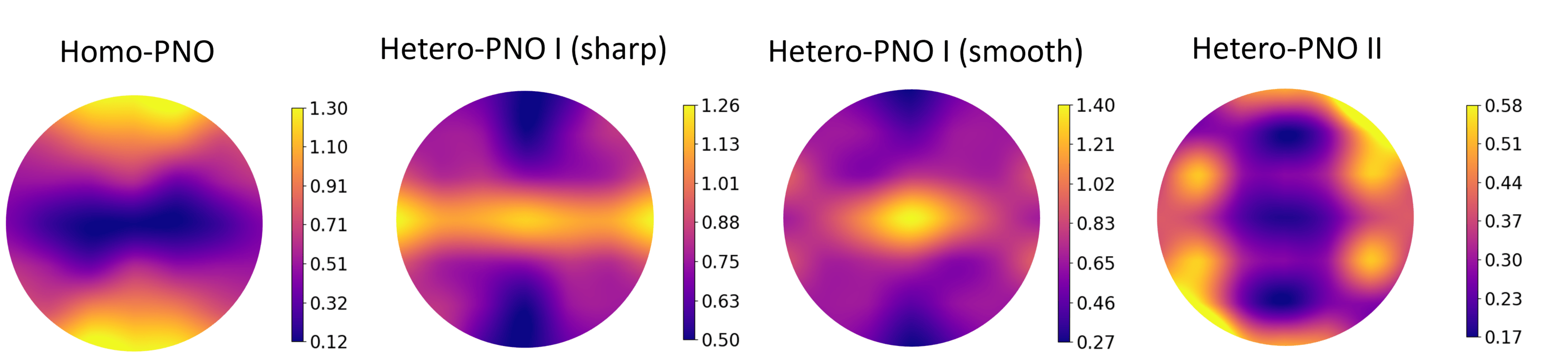}
 \caption{\YY{Synthetic dataset example 1: Learned influence (kernel) states for HomoPNO and the HeteroPNO models.}}
 \label{fig:HGO_kernels}
\end{figure}

\begin{figure}[!t]\centering
\includegraphics[width=.7\linewidth]{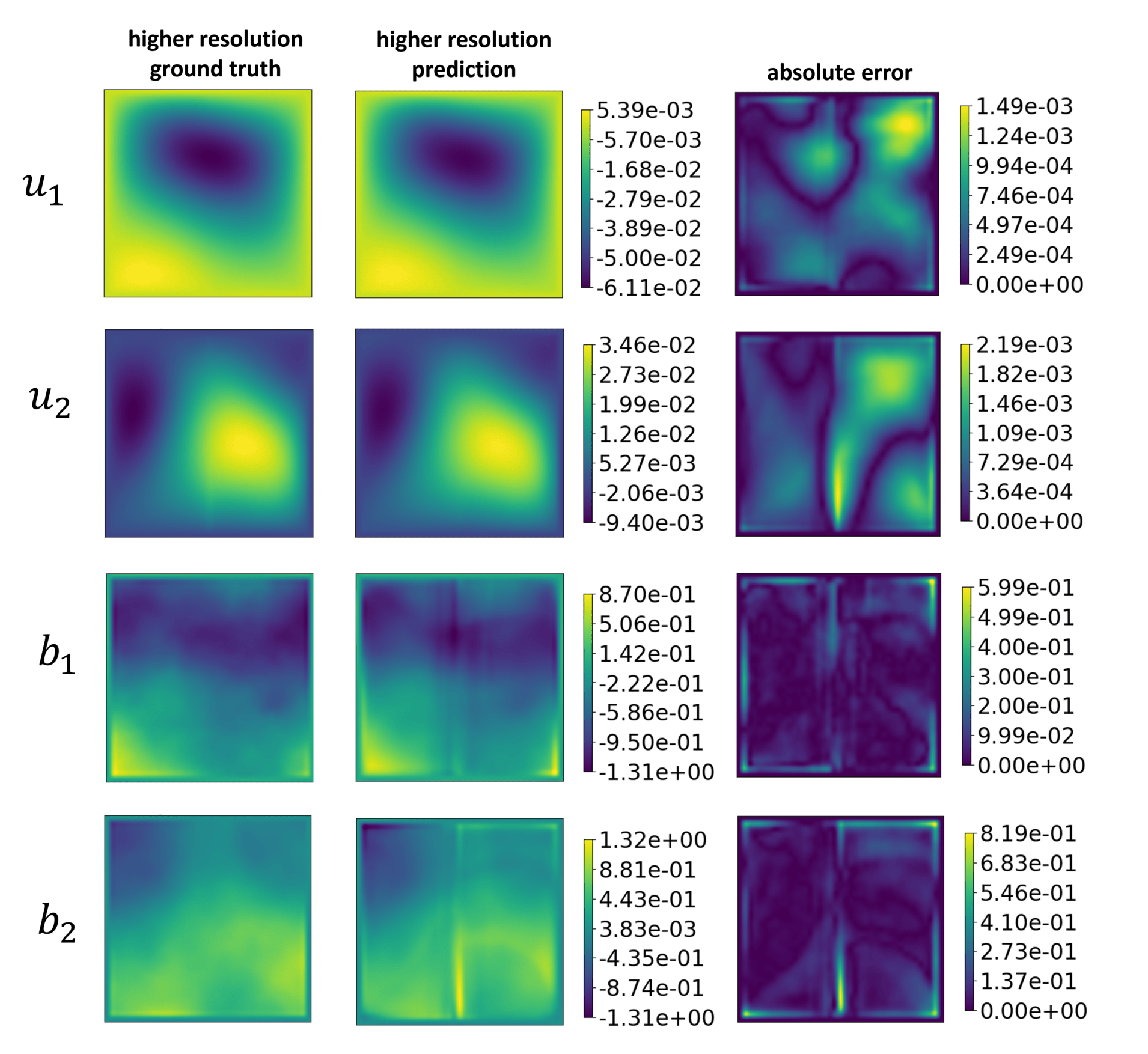}
 \caption{\YY{Synthetic dataset example 1: zero-shot predictions of a coarse-trained HeteroPNO II (on $21\times21$ grid) on a higher resolution grid ($41\times41$) against the high resolution ground truth.}}
 \label{fig:HGO_cross_res}
\end{figure}

We randomly split the 250 generated samples into training, validation, and test sets of size 200, 25, and 25, respectively. The widths of the MLPs for $\omega^{NN}$, $\sigma^{NN}$\YY{ and $\alpha^{NN}$} are (2, 256, 512, 1), (4, 512, 512, 1), \YY{and (2, 128, 128, 1)} respectively, and the peridynamic horizon size is set as $\delta = 3\Delta x$. In this example, since the data contains nonzero external forces, the setting falls under the category of Case I where \eqref{eqn: caseI_loss} is minimized via Algorithm \ref{alg:ML}. \YY{Table \ref{table:HGO_hyperparam} lists the hyperparameters tested with the Adam optimizer in PyTorch to train all PNO models, among which we pick the best model according to the separate validation dataset. In all experiments we decrease the learning rate with a ratio of decay rate every 100 epochs.}

\begin{figure}[!t]\centering
\includegraphics[width=.75\linewidth]{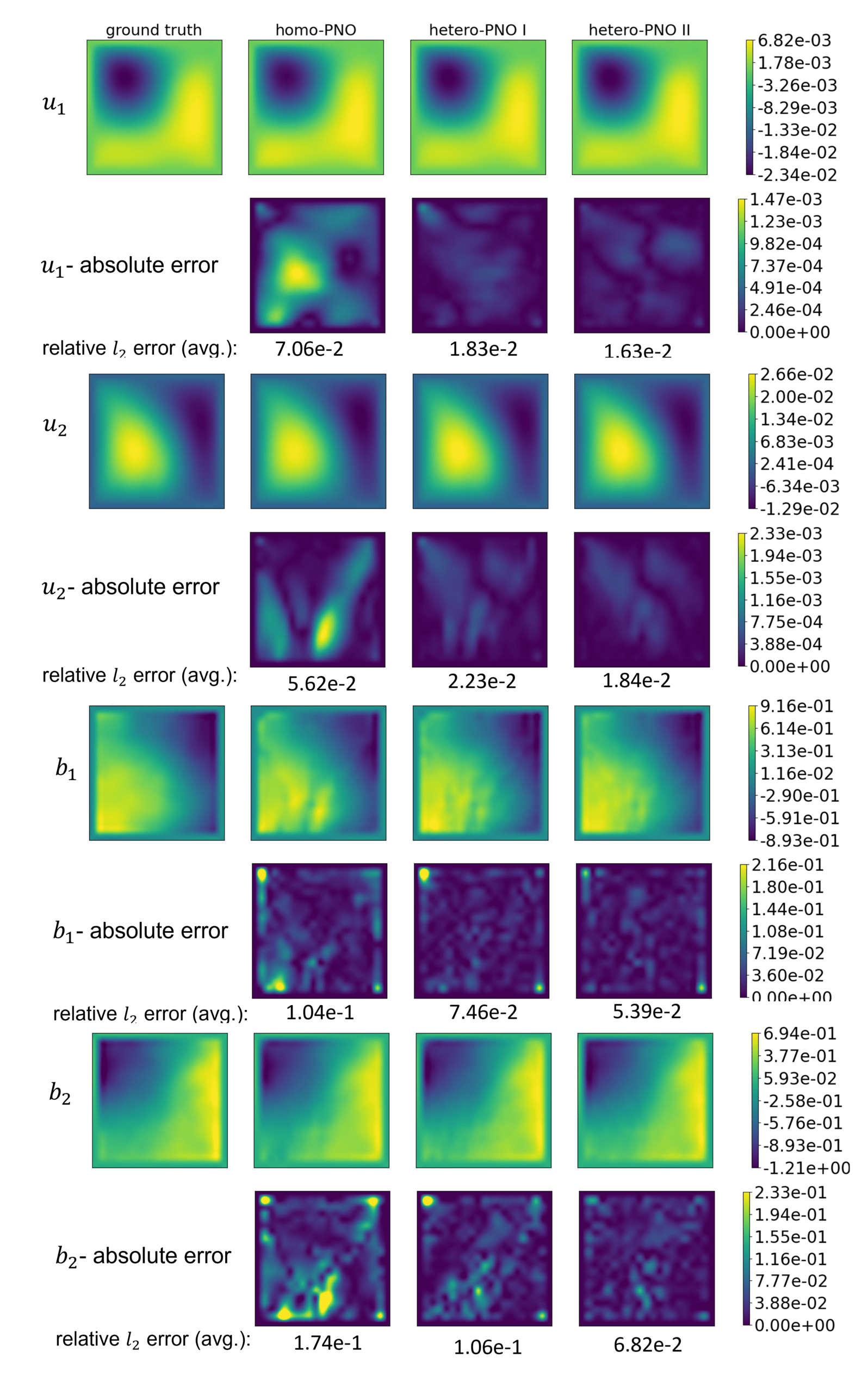}
 \caption{\YY{Synthetic dataset, example 2: HomoPNO and HeteroPNO predictions of displacement field (given external load and boundary conditions), and external forces (given displacement field) against the ground truth.}}
 \label{fig:HGO2_UB_results}
\end{figure}

\begin{figure}[!t]\centering
\includegraphics[width=.75\linewidth]{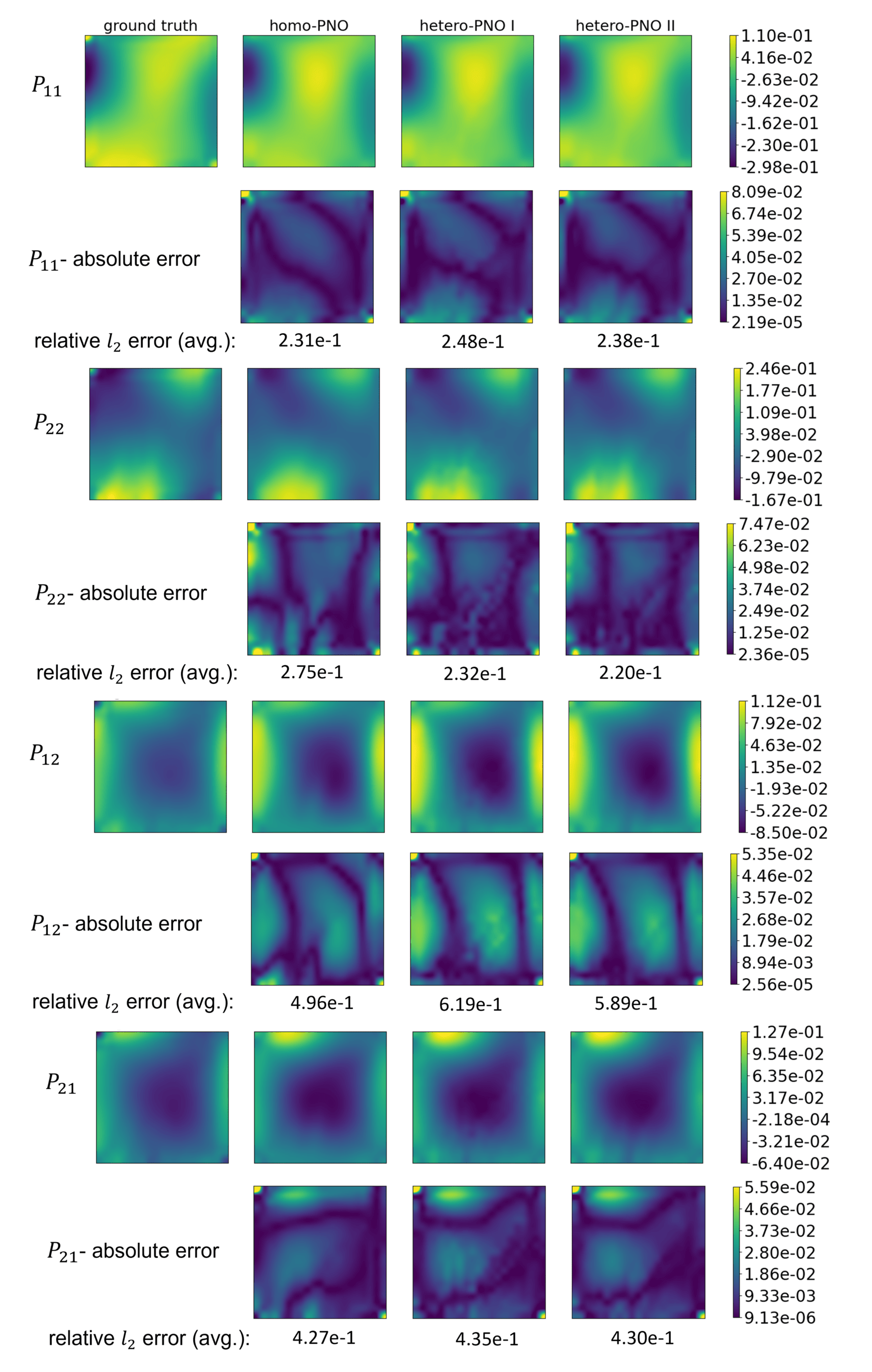}
 \caption{\YY{Synthetic dataset, example 2: HomoPNO and HeteroPNO predictions of the first Piola-Kirchhoff stress tensor against the ground truth.}}
 \label{fig:HGO2_stress}
\end{figure}

After training the PNO models, we report two different errors: 1) force error which reports the difference between predicted $-\mcG[\ub](\xb)$ and the applied external force density. 2) displacement error which reports the difference between the ground truth displacements and HeteroPNO predicted displacements using the static solver to solve for equilibrium under applied external force and Dirichlet boundary condition. These errors for the HomoPNO and HeteroPNO models are reported in \YY{Table \ref{table:HGO_errors} for example 1 and Table \ref{table:HGO2_errors} for example 2}. Here, HeteroPNO I and II, respectively, refer to scenarios \YY{where} microstructure/orientation field is considered as known input, or needs to be inferred from data. The predicted displacements and forces for one test sample are plotted in \YY{Figures \ref{fig:HGO_UB_results}, and \ref{fig:HGO2_UB_results} for the two examples}. We also plot the HeteroPNO predicted 1st Piola-Kirchhoff stress components versus the ground truth for the same test sample in \YY{Figures \ref{fig:HGO_stress} and \ref{fig:HGO2_stress}}. 

\YY{\textbf{Results of Synthetic dataset, example 1:} In this example, a simple orientation pattern with a sharp transition is modeled with the ``FNM'' boundary setting. From Table \ref{table:HGO_errors} we observe that the HeteroPNO architecture improves the force predictions by $50\%$ and the displacement prediction by $75\%$ compared to the HomoPNO. The HeteroPNO II model has slightly lower error compared to the HeteroPNO I with ground-truth $\alpha$. 
This is due to the fact that in this example, the ground-truth $\alpha$-map contains a sharp transition in the orientation field, which may not be exactly captured via a nonlocal model like PNO. 
One can see from the error plot in Figure \ref{fig:HGO_UB_results} that the heteroPNO I has noticeable errors around the transition region. HeteroPNO II, however, is unaware of the physical $\alpha$ near sharp transitions. Therefore, as one tries to fit a nonlocal model to such local data, an orientation field may be learned that slightly differs from the actual values but leads to lower error in predictions of material response. To further clarify this phenomenon, we have added an additional study, denoted as ``HeteroPNO I (smoothed $\alpha$)'', in Table \ref{table:HGO_errors}. In this study, we consider a smoothed version of the ground-truth fiber orientation $\alpha$. One can see that a smaller error is obtained.}
\YY{Additionally, we compare the predicted fiber orientations with the ground truth.} Figure \ref{fig:HGO_fiber} shows the predicted fiber orientation field by HeteroPNO II versus the ground-truth fibers assumed during data generation. \YY{The left is the ground-truth (with a sharp transition at center), the middle is the learned $\alpha$, and the right is the smoothed ground-truth which was used in a Hetero-PNO I model to study the influence of sharp versus smooth transitions on model's accuracy.}
\YY{This study also verifies that PNO learns a smoother transition between the left and right subdomains.} In Figure \ref{fig:HGO_kernels}, we plot the learned influence states for HomoPNO, HeteroPNO I and II. The kernel in HomoPNO with higher values along the vertical direction suggest that the material has a stiffer response in 90$^{\text{o}}$ which is consistent with the averaged fiber orientation assumed in the generated data. The base influence states of HeteroPNO I and II which are corresponding to $\alpha=0^{\text{o}}$ fiber directions, both show stiffer material response in horizontal axis, which is expected. Note that these influence states are then rotated at each location by depending on $\alpha$ at that location to align with local fibers.

\YY{The proposed PNO is an integral neural operator, and hence it should possess consistency across different grid resolutions. In other words, it can be used to predict fields on spatial resolutions which are different from the training data. To demonstrate this feature, we test the Hetero-PNO II which is trained on 21 $\times$ 21 grid, on a test sample with the higher resolution of 41 $\times$ 41. Figure \ref{fig:HGO_cross_res} shows the PNO predictions of forces and displacements versus the ground-truth high resolution data.}

\begin{table}[]
\begin{tabular}{|c|c|c|c|c|}
\hline
\multicolumn{2}{|c|}{error} & training (200) & validation (25) & test(25) \\
\hline
\multirow{4}{*}{force} & HomoPNO & {14.9}\% & {14.9}\% & {17.8}\% \\
 & HeteroPNO I & {7.80}\% & {8.05}\% &{9.15}\% \\
  & HeteroPNO II (FNM)& {5.62}\% & {5.77}\% &{8.97}\% \\
  & HeteroPNO II (FNM + $\alpha_{BC}$)& {6.14}\% & {6.45}\% &{9.09}\% \\
  & HeteroPNO II (VC)& {4.22}\% & {4.30}\% &{3.78}\% \\
\hline
\multirow{4}{*}{displacement} & HomoPNO & {5.83}\% & {5.76}\% & {2.41}\% \\
 & HeteroPNO I & {1.63}\% & {1.78}\% &{0.82}\% \\
  & HeteroPNO II (FNM)& {1.17}\% & {1.30}\% &{0.57}\% \\
  & HeteroPNO II (FNM + $\alpha_{BC}$)& {1.30}\% & {1.37}\% &{0.71}\% \\
  & HeteroPNO II (VC)& {1.04}\% & {0.84}\% &{0.70}\% \\
\hline
\end{tabular}
\caption{\YY{Synthetic dataset, example 2: Averaged relative $L^2$-norm error for HomoPNO and HeteroPNO predictions on forces (given displacement) and on displacement (given boundary conditions).}}
\label{table:HGO2_errors}
\end{table}



\begin{figure}[!t]\centering
\includegraphics[width=.89\linewidth]{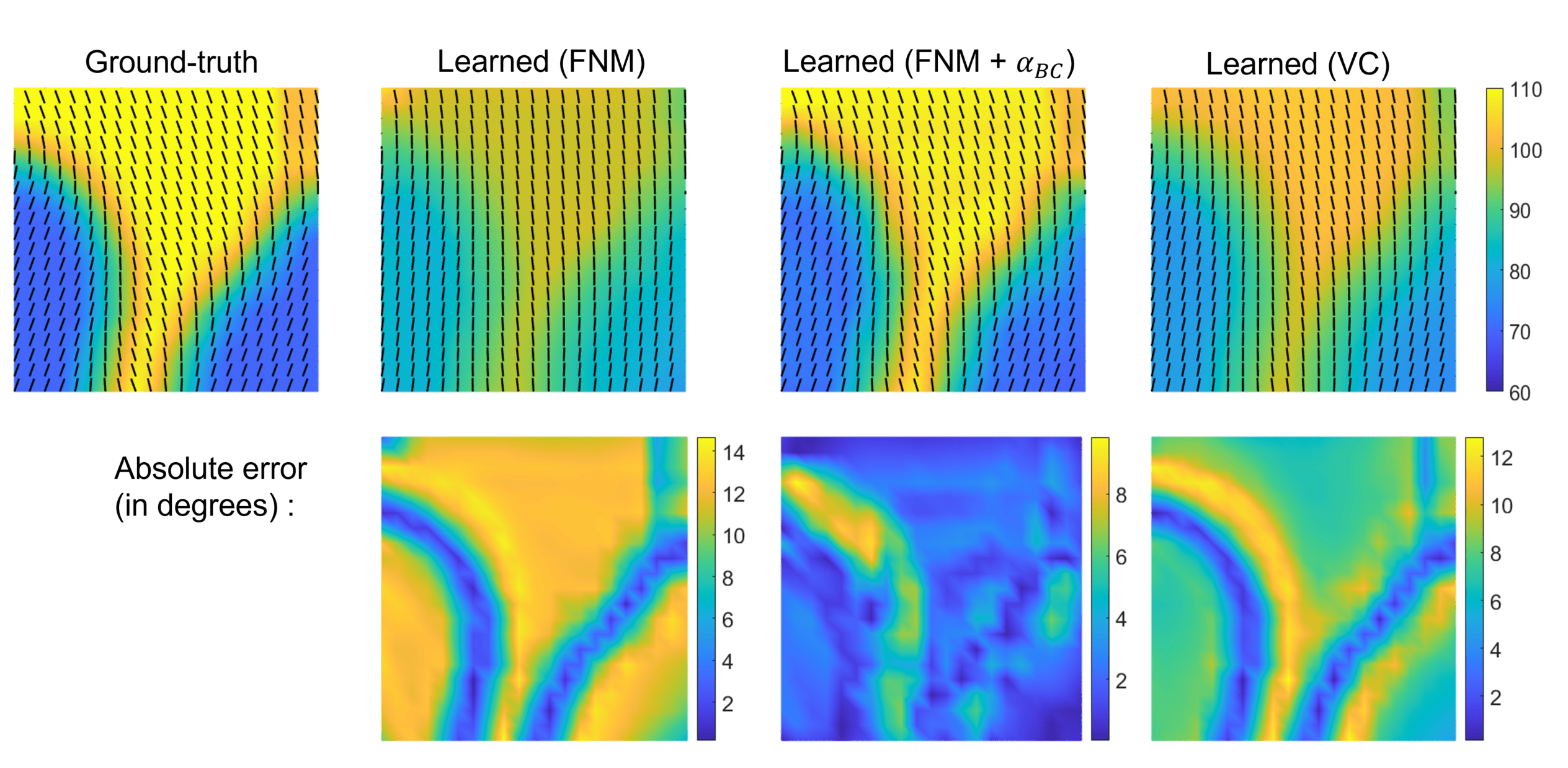}
 \caption{\YY{Synthetic dataset, example 2: Discovered hidden fiber orientation by Hetero-PNO models on the synthetic data set and comparison against the ground truth. The averaged absolute error for the learned orientation map for three boundary approaches FNM, FNM + $\alpha_{BC}$, and VC are respectively: 9.85$^{\text{o}}$, 2.67$^{\text{o}}$, 7.25$^{\text{o}}$.}}
 \label{fig:HGO2_fiber}
\end{figure}

\begin{figure}[!t]\centering
\includegraphics[width=.59\linewidth]{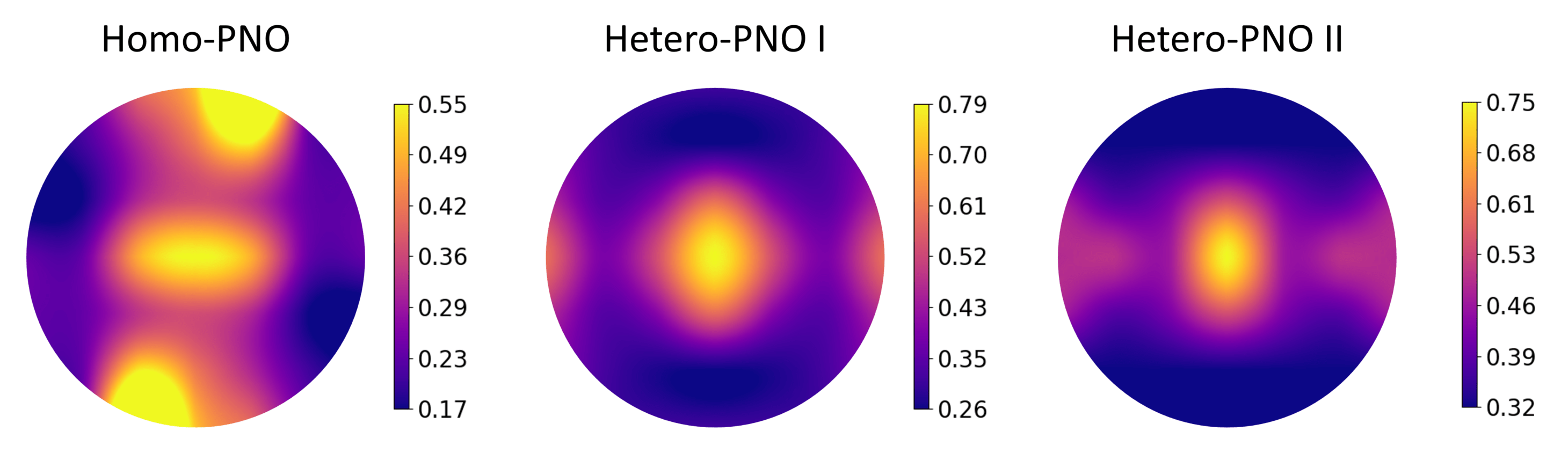}
 \caption{\YY{Synthetic dataset example 2: Learned influence (kernel) states for HomoPNO and the HeteroPNO models.}}
 \label{fig:HGO2_kernels}
\end{figure}

\YY{\textbf{Results of Synthetic dataset, example 2:} In this example, we consider a more complex fiber orientation pattern. To further clarify the effects associated with boundary conditions and provide an ablation study, here we additionally investigate different types of boundary settings. Three scenarios are consider:
\begin{enumerate}
\item ``FNM'': Set the learning domain as $\Omega=\Omega_{data}$, then extrapolate the displacement field to obtain the datum in $\Omega_I$ following the mirror-based fictitious nodes method \cite{zhao2020algorithm}. Infer $\vb$, $\wb$ together with $\alpha$ using Dirichlet-type boundary condition of displacement field on $\Omega_I$ and no boundary condition for $\alpha$. Notice that the data loss in \eqref{eqn: caseI_loss} is independent with the inferred fiber orientation, $\alpha(\xb)$, outside $\omg\cup \tilde{\omg}_I=\{\xb|\text{dist}(\xb,\omg)\leq \delta\}$. Hence, in this scenario we only infer $\alpha(\xb)$ on  $\omg\cup \tilde{\omg}_I$.
\item ``FNM+$\alpha_{BC}$'': Set the learning domain and perform displacement field extrapolation following the ``FNM'' approach. Infer $\vb$, $\wb$ together with $\alpha$ using Dirichlet-type boundary condition of displacement field on $\Omega_I$, together with an imposition of the ground-truth fiber orientation on $\tilde{\Omega}_I$ as a penalization term, for which the (modified) loss function in \eqref{eqn: caseI_loss} is defined as:
\begin{equation}\label{eqn: lossAlpha_caseI}
    \text{loss}_b = (1 - \gamma)\left(\frac{1}{S_{tr}}\sum_{s = 1}^{S_{tr}}\dfrac{\vertii{\mcG[\ub^s]+\bb^s}_{l^2(\omg)}}{\vertii{\bb^s}_{l^2(\omg)}}\right) + \gamma \frac{\vertii{\alpha - \alpha_{BC}}_{l^2(\tilde{\omg}_I)}}{\vertii{\alpha_{BC}}_{l^2(\tilde{\omg}_I)}} \text{ .} 
\end{equation}
where $\gamma$ is a tunable hyperparameter and $\alpha_{BC}(\xb)$ is the measured fiber orientation on $\tilde{\omg}_I$.
\item ``VC'': Given measurements on a larger domain: $\Omega \cup \Omega_I$, we can then use displacement values on $\Omega_I$ as a true volume constraint (VC) for PNO training on $\Omega$ instead of the fictitious nodes method. 
This approach avoids the possible additional errors from the FNM, but it would potentially waste some measurements on prescribing values on the boundary region. One learns $\vb$, $\wb$ together with $\alpha$ using Dirichlet-type boundary condition of displacement field on $\Omega_I$ and no boundary condition for $\alpha$. Herein, we again only infer $\alpha(\xb)$ in  $\omg\cup \tilde{\omg}_I$. 
\end{enumerate}
} 

\begin{figure}[!t]\centering
\includegraphics[width=.89\linewidth]{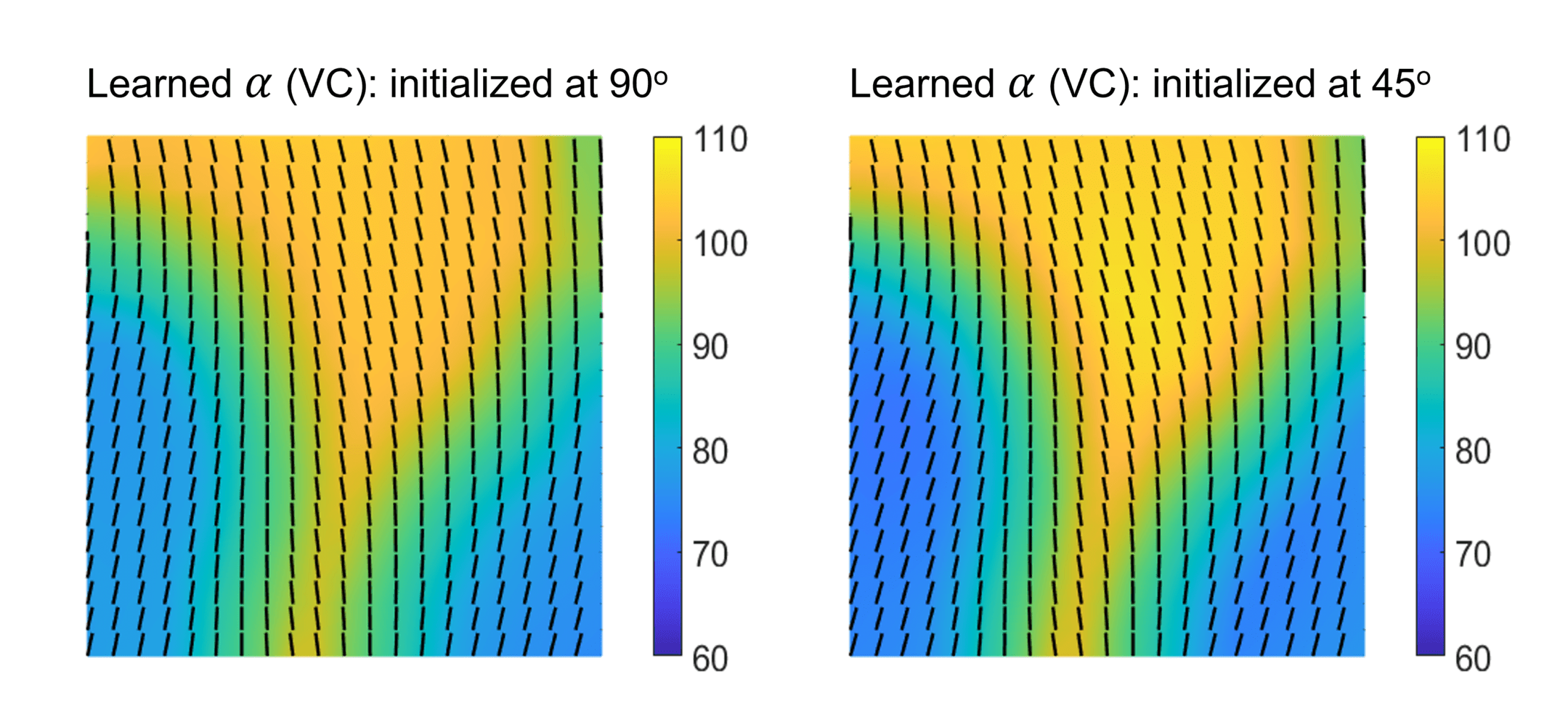}
 \caption{\YY{Synthetic dataset, example 2: initialization independence, when learning orientation map using Hetero-PNO II (VC), with two different initialization of $\alpha$. The average absolute errors for initialization at $90^{\text{o}}$ and $45^{\text{o}}$ are respectively: $7.25^{\text{o}}$ and $5.48^{\text{o}}$.}}
 \label{fig:alpha_init}
\end{figure}

\begin{figure}[!t]\centering
\includegraphics[width=.7\linewidth]{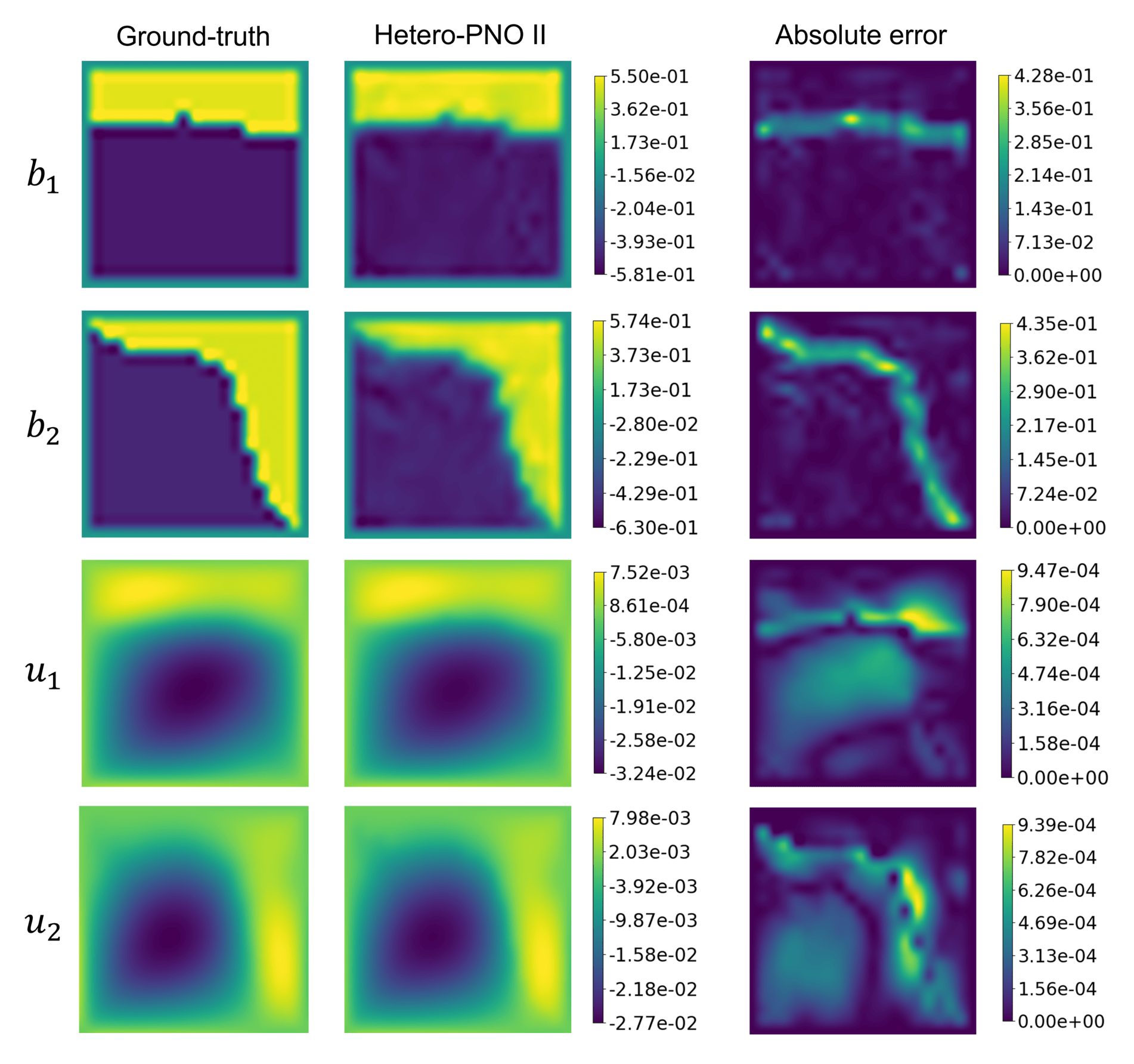}
 \caption{\YY{Synthetic dataset, example 2: predictions of the trained HeteroPNO II (FNM + $\alpha_{BC}$) model on a specific test data generated via discontinuous external forces.}}
 \label{fig:HGO_discont}
\end{figure}


\YY{Table \ref{table:HGO2_errors} demonstrates the errors of displacement and force fields, and Figure \ref{fig:HGO2_fiber} shows the learned orientation fields from these three approaches. One can observe that the ``VC'' setting has the smallest error in displacement and force fields, while the ``FNM+$\alpha_{BC}$'' setting has comparable errors in these fields. Moreover, while all three approaches lead to acceptable discovered microstructure, the ``VC'' and ``FNM+$\alpha_{BC}$'' lead to a more accurate orientation field. This investigation verifies the fact that the boundary treatment such as ``FNM'' used in a heterogeneous setting would introduce errors in fiber orientation, and these errors can be reduced either by using the ground-truth displacement measurements in the nonlocal boundary region (as in ``VC'' setting) or introducing a boundary condition for the fiber orientation field (as in the ``FNM+$\alpha_{BC}$'' setting). } 

\YY{Additionally, with this example we study the influence of $\alpha$ initialization on the learned microstructure and concluded that Hetero-PNO II is robust with respect to various initialization. The learned $\alpha$ with two different initialization are plotted in Figure \ref{fig:alpha_init}: both are consistent with the ground-truth fiber orientation in this case.}

\YY{Lastly, since PNO is a constitutive model, it should be agnostic to loading patterns, domain shapes, and boundary conditions. While generalization to unseen domain shapes and boundary conditions has been demonstrated in \cite{jafarzadeh2024peridynamic}, here we show the generalization to unseen loading patterns. Specifically, we test the Hetero-PNO II model trained on smooth random external loadings on a case under discontinuous loading. Figure \ref{fig:HGO_discont} shows the PNO's predictions against the ground-truth for this scenario. Consistent prediction is again obtained.}

\section{Application on DIC measurements of Bio-tissues}\label{sec:tissue}

Having illustrated the performances of our learned neural operators on high-fidelity synthetic simulation datasets, we now consider a problem of learning the material response of a tissue sample from DIC displacement tracking measurements as a prototypical exemplar. The main objective of this section is to provide a proof-of-principle demonstration that the framework introduced thus far applies to discover the constitutive equations and material microstructure and to estimate the stress field, while the dataset has unavoidable measurement noise. In this application we further compare our proposed HeteroPNO against two conventional approaches that use constitutive modeling with parameter fitting to demonstrate the advantages of neural operator models and the importance of considering the heterogeneity of material microstructures.

\subsection{Data Collection and Preparation}

\begin{figure}[!t]\centering
\includegraphics[width=.99\linewidth]{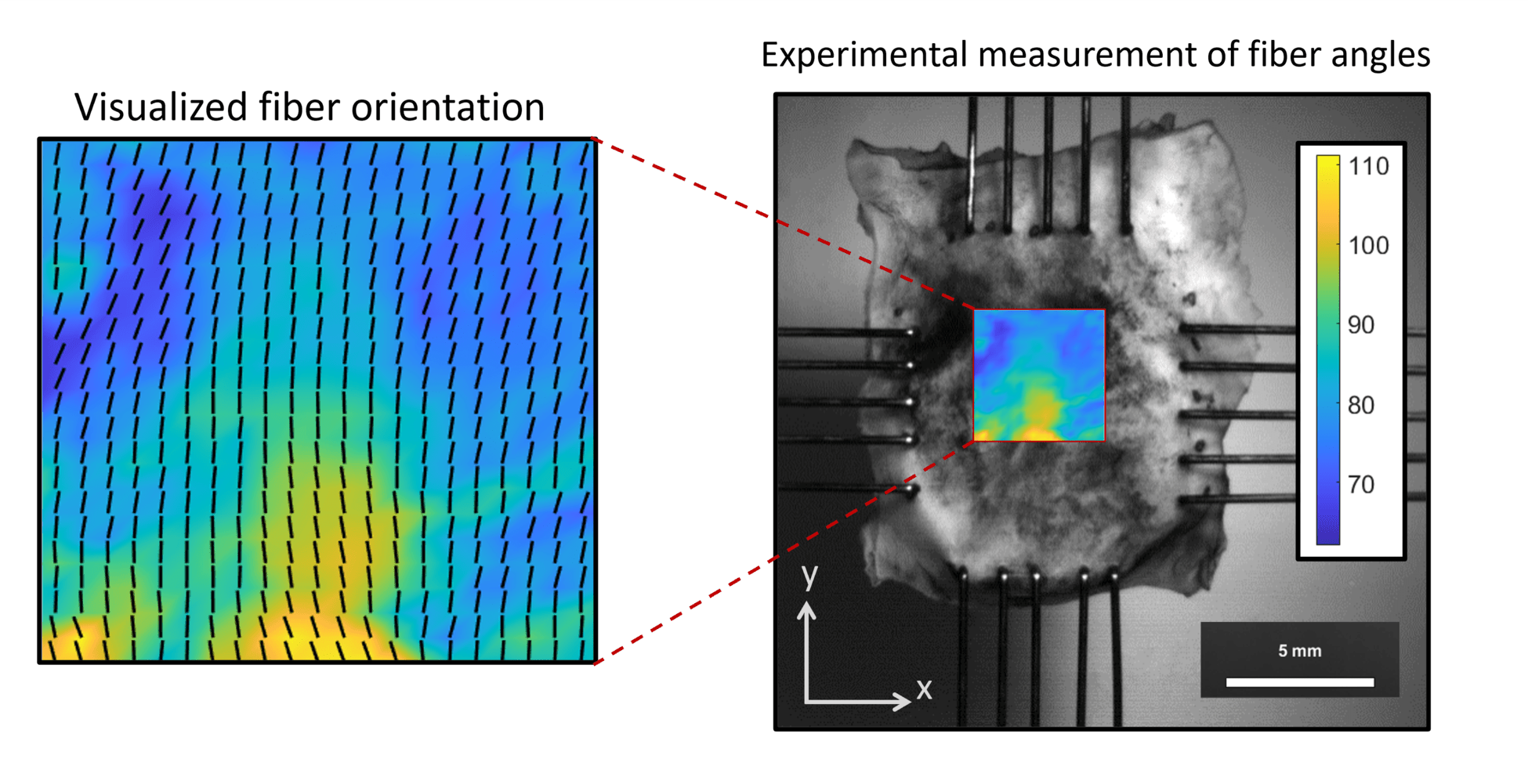}
 \caption{Experimental collagen fiber orientation field.}
 \label{fig:fiber_exp}
\end{figure}

\begin{figure}[!t]\centering
\includegraphics[width=.79\linewidth]{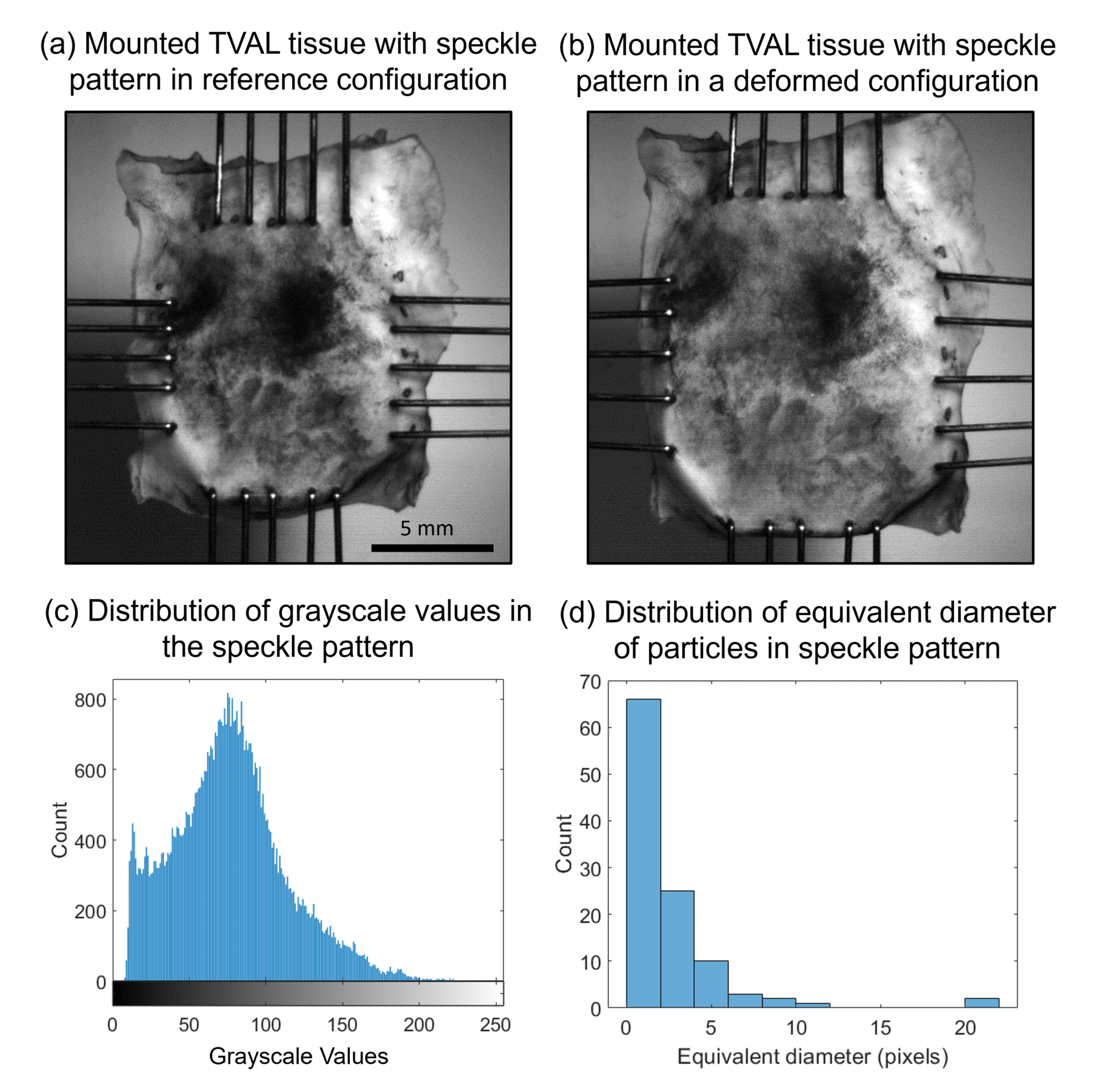}
 \caption{\YY{Mounted TVAL tissue with speckle pattern applied on the surface for digital image correlation (DIC) measurement of displacement field.}}
 \label{fig:DIC_speckle}
\end{figure}

We have acquired biaxial mechanical testing data and collagen fiber microstructure from a representative porcine tricuspid valve anterior leaflet (TVAL) tissue, following our established experimental procedures \cite{jett2018investigation,jett2020integration, laurence2022benchtop}. In brief, one adult porcine heart ($n=1$, 120 kg, 1.5 years of age) was obtained from a USDA-approved abattoir (Animal Technologies, Inc., Tyler, TX, USA) within a day of animal slaughtering. Upon arrival at our laboratory, we then sectioned the TVAL tissue into a square specimen and measured the thickness at three locations using an optical measuring system (Keyence, Itasca, IL, USA) to obtain the average tissue thickness ($L_z=0.22 mm$). We next introduced a speckling pattern to the tissue surface using an airbrush and black paint \cite{you2022physics}, that was later used in digital image correlation (DIC)-based tracking of tissue displacement/deformation. \YY{We have utilized 4-9 fiducial markers to confirm that the applied speckle pattern could accurately track the displacement field.}
The painted specimen was then mounted to a commercial biaxial testing system (BioTester, CellScale, Waterloo, ON, Canada), resulting in an effective testing area of $L_x \times L_y = 8.72 \times 10.75 mm$ for the subsequent tissue biaxial mechanical characterizations. \YY{The experimental photos of the speckle pattern applied to the tissue surface at the undeformed (left) and loaded (right) configurations are provided in the top row of Figure \ref{fig:DIC_speckle}), where the speckle pattern quality were evaluated using three criteria: (i) range of grayscale values, (ii) the particle size, and (iii) the global mean intensity gradient (MIG) metric. From the histogram on the bottom left of Figure \ref{fig:DIC_speckle}, one can see that a relatively wide spectrum of grayscale values that indicates no underexposure or overexposure. Then, we further examined the equivalent diameter, which is defined as the diameter of a sphere that has the same area as the particle, i.e.  $D_{equiv}= (4A/\pi)1/2$ where $A$ is the area of the particle. We found that most particles in the speckle pattern have an equivalent diameter of 1-5 pixels (the bottom-right histogram), and the mean diameter is 5.69 pixels that is within the desirable 3–7 pixel range according to \cite{sutton2008digital}. Finally, we calculated the mean intensity gradient (MIG) proposed by Pan \textit{et al.} \cite{pan2008study}, which is a commonly adopted global parameter to assess the quality of the pattern. The MIG value for the speckle pattern in this experiment is 8.12 which may be at a lower end of desirable speckle pattern (i.e., MIG$>$10 as suggested by Pan \textit{et al.}). Despite this less desired MIG value for the applied speckle pattern, we expect the experimentally measured displacement and strain fields would be adequate to serve as the data for calibrating and validating of our proposed PNO method with application to modeling the biaxial mechanical behavior of planar biological tissues.}

\begin{figure}[!t]\centering
\includegraphics[width=.79\linewidth]{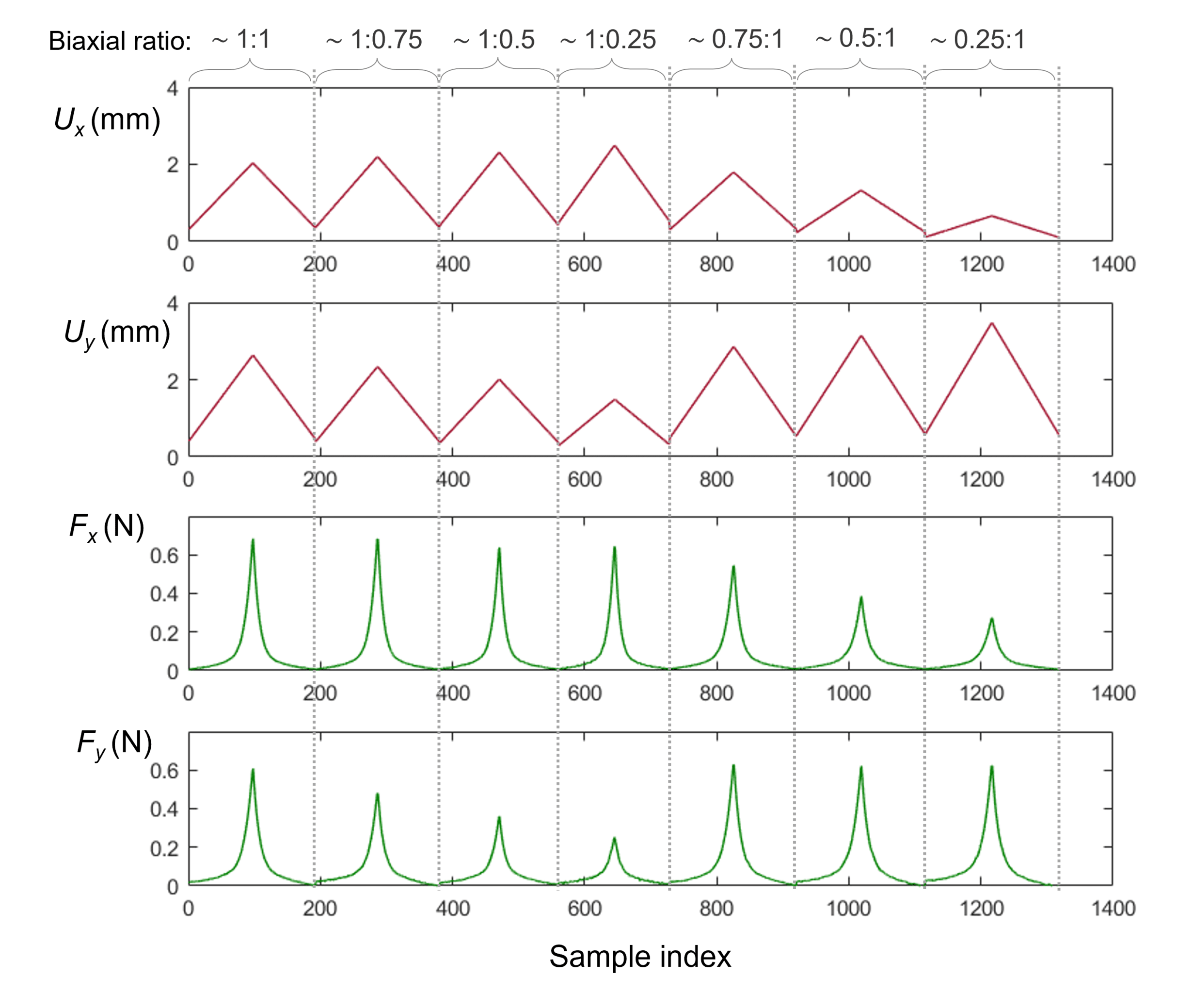}
 \caption{\YY{Displacements and forces of the tines in the biaxial test versus sample index.}}
 \label{fig:UF_biaxial}
\end{figure}

First of all, the sample was immersed in a phosphate-buffered saline (PBS) and underwent a preconditioning protocol that consists of 10 cycles of equi-biaxial tension loading and unloading, targeting a first Piola-Kirchhoff stress of 150 kPa for emulating the in vivo functioning conditions of the tricuspid valve \cite{jett2018investigation}. Next, seven displacement-controlled biaxial tension protocols were performed, considering the following biaxial stresses: $P_{11}:P_{22}=1:1,1:0.75,1:0.5,1:0.25,0.75:1,0.5:1,$ and $0.25:1$. \YY{The tine displacement and forces are plotted in Figure \ref{fig:UF_biaxial} versus temporally-sequenced samples.}
Here, $P_{11}=f_x⁄{L_xL_z}$  and $P_{22}=f_y{⁄L_yL_z}$  are the 1-1 and 2-2 first Piola-Kirchhoff (1st PK) stress components, respectively. Since the specimen was mounted with tissue’s circumferential direction aligned with the $x-$direction of the BioTester, they can also be viewed as the 1st Piola-Kirchhoff stress components in the $x-$ and $y-$directions. The corresponding stretches can also be computed as the ratio of the deformed to the undeformed tine distances. Each of the above seven stress ratios was performed for five loading/unloading cycles. From the last load and unloading cycle of each stress ratio, $1280\times960$ resolution images of the specimen were captured by a CCD camera, and the load cell readings ($f_x$ and $f_y$) were recorded at 5 Hz, resulting in 1,318 data points.

\YY{Note that the rigid-type BioRakes hooks in mounting the tissue specimen may cause the local stress concentrations near the tines/tips of the hooks, which agrees with what we also found from a numerical study of biaxial testing \cite{laurence2023investigation}: (i) more profound stress concentrations occur near the hooks, (ii) but the stress becomes relatively more homogeneous for the central one-third ~ one-half of the tissue specimen. To minimize such boundary loading effects, we followed the common practice in soft tissue experimental biomechanical testing and analysis to determine the mechanical behavior for the smaller region of interest (ROI) of the tissue specimen: i.e., in this study we chose the central one-half (4.4$\times$4.4 mm) region for the deformation analysis, where the effective size of testing area was 8.72$\times$10.75 mm which is the area delimited by the hooks/tines. On the other hand, the force sensor (load cell) comes with a 0.1\% resolution (i.e., 0.0015 N or 1.5 mN) which yielded fairly smooth force-displacement for the raw experimental data shown in Figure \ref{fig:UF_biaxial}.}

\YY{We also note that tissues can present viscoelastic behaviors. However, in our displacement-controlled biaxial testing, the displacement rate was approximately 0.16 mm/s in the X-direction and 0.28 mm/s in the Y-direction. That is equivalent to a stretch rate of $\sim 0.02\text{ s}^{-1}$ and $0.04\text{ s}^{-1}$ in the two directions, respectively, which is slow enough to capture the quasi-static behaviors (i.e., the viscous or time-dependent effect is nearly negligible) \cite{carew2000role,robinson2009planar,talman1995glutaraldehyde,noble2020rate,stella2007time}. In addition, in our previous experimental studies for mechanical behaviors of cardiac heart valves, we found that with this adopted displacement rate (or mimicking the quasi-elastic regime) we were able to recover repeated mechanical characterization results and the mechanical responses obtained from the loading phase were very close to those from the unloading phase (i.e., very minimum hysteresis). Therefore, we focus on a quasi-static setting in this study.}

\paragraph{\textbf{Collagen Microstructural Imaging.}}

\begin{figure}[!t]\centering
\includegraphics[width=.79\linewidth]{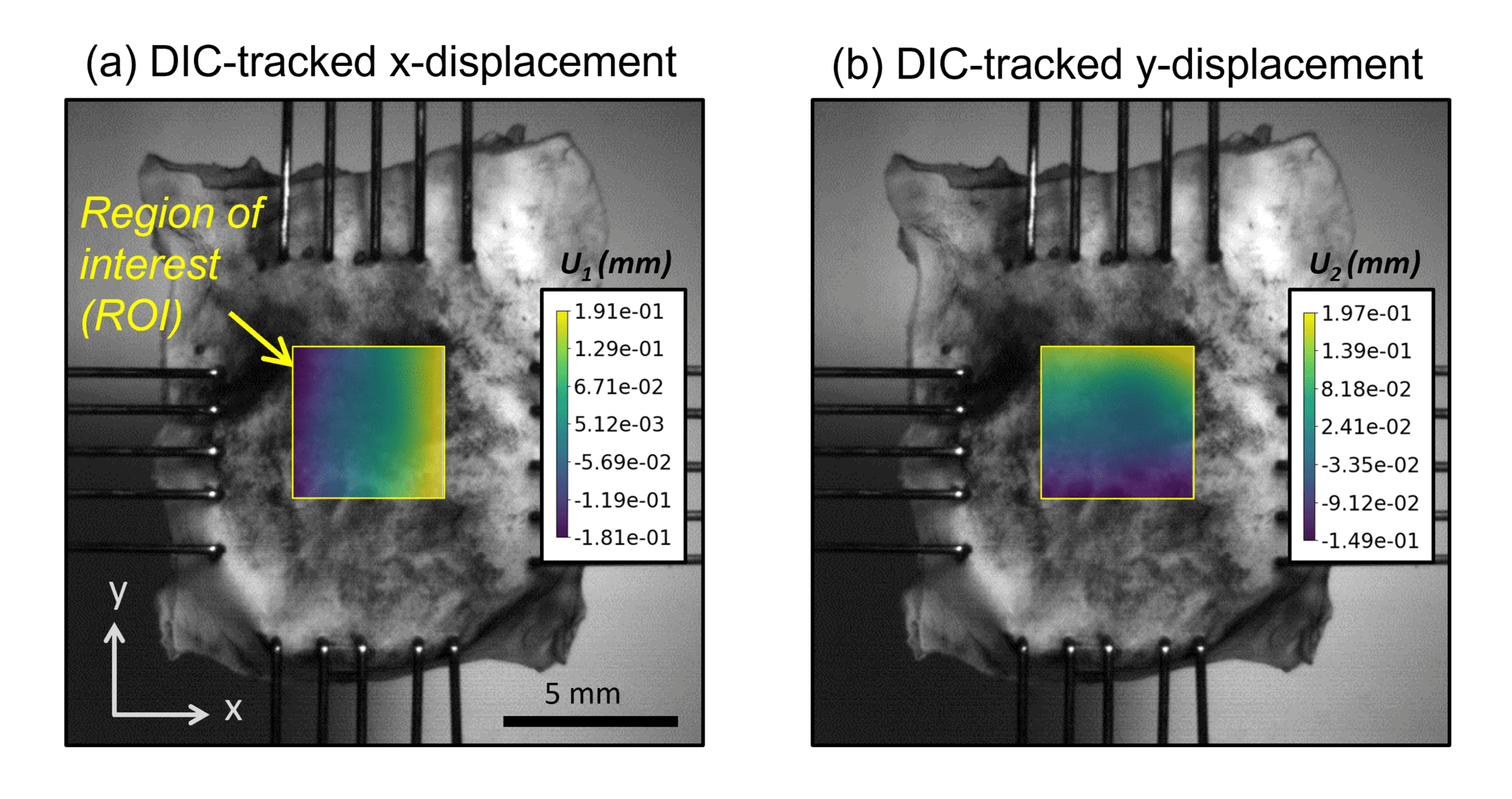}
 \caption{\YY{Demonstration of displacement field measurements via digital image correlation (DIC) on the region of interest.}}
 \label{fig:DIC_exp}
\end{figure}

After mechanical testing, we further performed collagen microstructural imaging via an in-house polarized spatial frequency domain imaging (pSFDI) device to obtain the collagen fiber orientation map of the tissue specimen. Following the procedure of pSFDI-based collagen fiber quantifications, the incident spatial frequency light patterns were produced from an LED projector (Texas Instruments, Dallas, TX) with a wavelength of 490 nm (cyan). A 1.2-megapixel CCD camera (Basler, Germany) was used to capture the reflected light intensity responses through a rotating linear polarizer (Thorlabs Inc., Newton, NJ) at 37 distinct polarization states (i.e., 0$^\circ$ to 180$^\circ$, 5$^\circ$ increments). This imaging procedure was repeated for three linear phase shifts (0$^\circ$, 120$^\circ$, and 240$^\circ$) of the spatial frequency pattern. Image processing and data analyses were then completed via custom MATLAB (MathWorks, Natick, MA) programs to examine the pixel-by-pixel collagen fiber orientation ($\alpha$) of the tissue’s region of interest (ROI) at the post-preconditioning state that was chosen as the reference (undeformed) configuration. Please refer to more details of pSFDI data analysis in Goth \textit{et al.} \cite{goth2019non} and Jett \textit{et al.} \cite{jett2020integration}. The fiber orientation field is further smoothed using convolution with a square pulse of size $21\time21$ pixels to reduce measurement noises.Here, the central $4.4\times4.4$ $mm^2$ region was selected as the tissue ROI for our collagen fiber microstructure analysis and DIC-based displacement tracking (Figures \ref{fig:fiber_exp} and \ref{fig:DIC_exp}). \YY{In Figure \ref{fig:fiber_exp} (right), the colormap that shows the degree of fiber orientation angle across the ROI with respect to the \textit{x}-axis is superimposed on the tissue sample photo. On the left is the magnified ROI colormap with dashed lines which represent those fiber angles on our 21x21 grid computational domain. Further details and other representative pSFDI measurements for bovine tendon and mitral valve posterior leaflet may be found in \cite{jett2020integration}.}


\paragraph{\textbf{Digital Image Correlation (DIC)-based Displacement Tracking.}}

An open-source DIC software Ncorr \cite{blaber2015ncorr} was used in this study to extract the full-field displacements from the recorded images. A rectangular ROI was considered in the reference image (the colored regions in Figure \ref{fig:DIC_exp}), which is partitioned into subsets of smaller regions. A subset size of 40 pixels (0.878 mm) and a subset spacing of 2 pixels (0.0439 mm) were selected according to DIC processing principles \cite{pan2008study} and the speckle patterns in this study. Each distinctively identified pair of corresponding subsets in the reference and deformed images are correlated based on the normalized cross-correlation (NCC) criteria \cite{blaber2015ncorr} (difference vector norm cutoff 1e-6, iteration count cutoff 100). Since the TVAL specimen was subjected to large deformations during the equi-biaxial experiments, the reference image was updated multiple times during the DIC analysis to continuously keep track of the positions of largely deformed material regions.  This was achieved by enabling the ‘high-strain analysis’, ‘seed propagation’, and ‘auto propagation’ functionalities in Ncorr. The full-field displacement components shown in Figure \ref{fig:DIC_exp} was obtained in the software by the following algorithm: 
\[\tilde{x}_i-x_i=u_0+\frac{\partial u}{\partial x}(x_i-x_0)+\frac{\partial u}{\partial y}(y_i-y_0)\]
\[\tilde{y}_i-y_i=v_0+\frac{\partial v}{\partial x}(x_i-x_0)+\frac{\partial v}{\partial y}(y_i-y_0)\]
where  $(x_0, y_0)$  and  $(u_0,v_0)$  are the coordinate and displacement components of a reference subset center\YY{, and} $(x_i,y_i)$  and  $(u,v)$ are coordinate and displacement components of an initial reference subset point. Coordinates of that reference subset point in the deformed state is denoted by $(\tilde{x}_i,\tilde{y}_i)$.
Afterwards the displacements in ROI was further smoothed by convolving with a square pulse of size $105\times105$ pixels to reduce the DIC processing artifacts. 

\YY{As pointed out by previous studies \cite{sugerman2023speckling,liu2019application}, DIC measurements have unavoidable errors but in general, if all the experimental parameters are controlled appropriately, the error in strain measurements is negligible. For example, Blenkinsopp \textit{et al.} \cite{blenkinsopp2019method} showed that the measurement has $<0.05\%$ error under large strain of $\sim 50\%$. Herein, we followed these methodologies in our present study to perform our DIC deformation and strain analysis and expect the strain measurements to have a similar level of accuracy (or comparable level of errors).}

\subsection{Learning material model and microstructure discovery}

The 1318 collected data is split into training, validation and test sets as follows. \YY{With the purpose of testing PNO in a small data regime, the training set used for this problem consists of 100 samples equi-spaced temporally over all cycles. This set includes about fourteen samples per loading/unloading cycle. Additionally, we also study the performance of PNO models with further reduced training samples (25 and 50 samples, respectively), to demonstrate how the number of training samples affect the models accuracy.
To generate a fair comparison among all experiments, a validation set consisting of 20 samples is selected from the remainder of the data, such that instances with high, low, and mid-level stretches from all seven protocols are included. The remainder 1098 samples are used as the test set to evaluate the accuracy of the models. } To have a base model, first, we calibrate a popular classical hyperelastic constitutive law using the training set. Then we train a HomoPNO, HeteroPNO I (assuming fiber orientation field are given) and HeteroPNO II (unknown fiber angles to be discovered) in a similar fashion to the \YY{previous} example.

\paragraph{\textbf{Baseline Constitutive Model.}}
For the baseline model, we use one of the popular classical constitutive laws, Fung model, which is frequently used for biological tissues. The Fung model here has the strain energy density of form:
\[\psi=\frac{c}{2}\left[exp(a_1E_{11}^2+a_2E_{22}^2+2a_3E_{11}E_{22})\right]\text{,}\]
where $c$, $a_1$, $a_2$, and $a_3$ are the model parameters to be determined, and $E_{11}$, $E_{22}$ are the principle Green–Lagrange strains in the $x-$ and
$y-$directions, respectively. Here we have used two approaches to determine the model parameters from data. One approach, we we call it \textit{forward} calibration, is to find the parameters by fitting the model predicted stretch-stress curves to those obtained from the DIC measurements. To this aim differential evolution optimization is employed
which minimizes the residual mean squared errors in the stress between the experimental data (training set only) and model prediction \cite{price2006differential}.

The second calibration approach which we refer to it as \textit{inverse}, is to define the optimization objective function on the finite-element-obtained displacement error instead of the stresses. This requires FE analysis for each optimization step on all training samples to provide nodal displacement predictions which is then passed onto the objective function to be compared with the DIC measurements. \YY{In particular, we used Bayesian global optimization routine with the objective (loss) function expressed as the square sum of the point-wise displacement errors (between the DIC measurements and finite element predictions) weighted together with the 1st PK stress discrepancy. As such, the optimization continuously explores new possible parameter set(s) in the parameter space during the iteration, making our calibration approach less sensitive to the initial guess. Further numerical investigations of the application of Bayesian optimization can be found in \cite{ross2024bayesian}.}

\YY{With both calibration approaches, we infer Fung models with homogeneous parameters, and then denote as ``Homogeneous Fung (forward)'' and ``Homogeneous Fung (inverse)'', respectively. Additionally, to provide a fair comparison with the heterogeneous PNO models, we also construct a Fung model with heterogeneous fiber orientations. In particular, the measured fiber orientation is substituted into the Fung model, with the rest of parameters treated as homogeneous and inferred using the forward approach. This case can be seen as an analog of the HeteroPNO I case, and is denoted as ``Heterogeneous Fung (forward)''.} Once the model parameters are \YY{obtained} from either approaches, they are used in Abaqus/standard solver \cite{abaqus2011abaqus} to predict the displacements for samples in the test set. The FEM setting in inverse calibration and also in the testing, is to solve for equilibrium using the DIC displacement on the domain edges as Dirichlet boundary conditions.

\YY{Besides learning a constitutive operator, as shown in previous works \cite{you2022physics,zhang2023metano} neural operator models \cite{li2020fourier,lu2019deeponet} can also be employed to model heterogeneous tissue responses as a mapping from boundary datum $\ub_{BC}(\xb)$ to the corresponding displacement field $\ub(\xb)$, as an implicit PDE solver. However, compared with these approaches, the proposed PNO approach learns an explicit constitutive law, making it possesses two advantages:
\begin{enumerate}    
\item PNO guarantees the balance of linear and angular momentum as well as material objectivity, due to its design of the architecture according to the ordinary-state-based peridynamics theory. This property makes PNO interpretable and physically consistent.
\item As a constitutive law, PNO enables the calculation of stress fields. As such, one can match not only the displacement fields but also the first Piola-Kirchhoff axial stresses from experimental measurements.
\end{enumerate}
On this example, we also construct a $4-$layer FNO model \cite{li2020fourier}, with a comparable number of trainable parameters (26.8k) and report its performance.}


\begin{table}[]
\begin{tabular}{|c|c|c|c|c|c|c|}
\hline
Model&\multicolumn{3}{c|}{HeteroPNO II}&\multicolumn{3}{c|}{FNO}\\
\hline
$N_{train}$ & training error & validation error & test error&training error & validation error & test error\\
\hline
25 & {4.40}\% & {10.45}\% & {7.74}\% & {5.82}\% & {9.45}\% & {6.02}\%\\
\hline
50 & {4.27}\% & {8.29}\% & {6.35}\% & {4.73}\% & {7.22}\% & {4.67}\%\\
\hline
100 & {3.47}\% & {7.92}\% & {5.27}\% &  {3.65}\% & {5.51}\% & {3.63}\%\\
\hline
\end{tabular}
\caption{\YY{Displacement errors using different number of training samples.}}
\label{table:tissue_ntrain}
\end{table}



\begin{table}[]
\begin{tabular}{|c|c|c|c|}
\hline
model & training(100) & validation(20) & test(1198)\\
\hline
{Homogeneous Fung (forward)} & 19.60\% & 23.60\% & 20.09\% \\
\hline
{Homogeneous Fung (inverse)} & 19.01\% & 22.74\% & 19.30\% \\
\hline
{Heterogeneous Fung (forward)} & 22.53\% & 27.29\% & 22.41\% \\
\hline
HomoPNO & 8.48\% & 12.43\% & 9.95\% \\
\hline
HeteroPNO I & 7.65\% & 11.42\% & 8.50\% \\
\hline
{HeterPNO II}& 3.47\% & 7.92\% & 5.28\% \\
\hline
{FNO}& 3.65\% & 5.51\% & 3.63\% \\
\hline
\end{tabular}
\caption{Averaged relative $l_2$-norm error of different models' prediction for the displacement field given boundary conditions.}
\label{table:tissue_errors}
\end{table}

\begin{figure}[!t]\centering
\includegraphics[width=.80\linewidth]{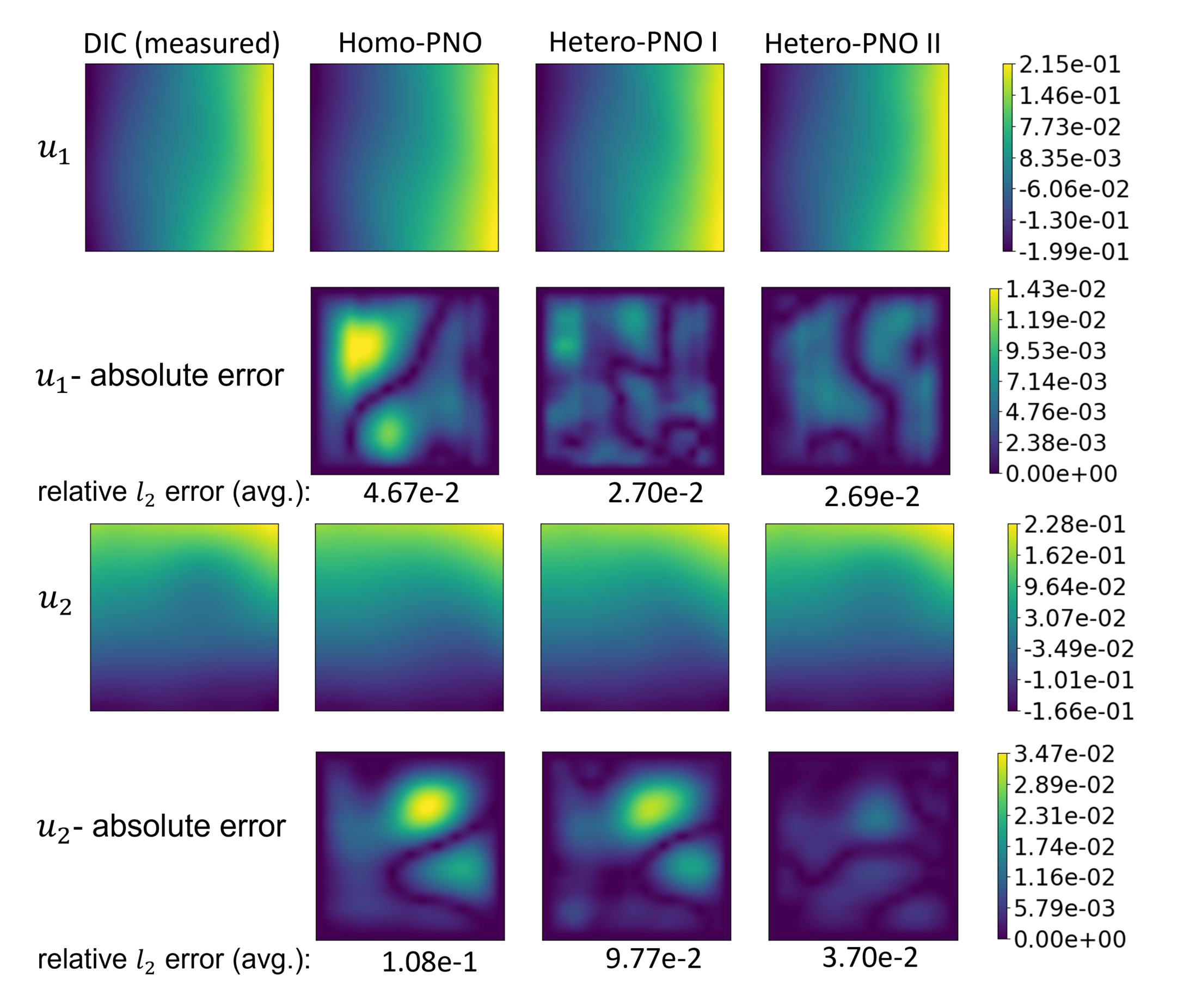}
 \caption{Predicted displacement field by HomoPNO and HeteroPNO models \YY{versus DIC measurements}.}
 \label{fig:tissue_disp}
\end{figure}

\begin{figure}[!t]\centering
\includegraphics[width=.89\linewidth]{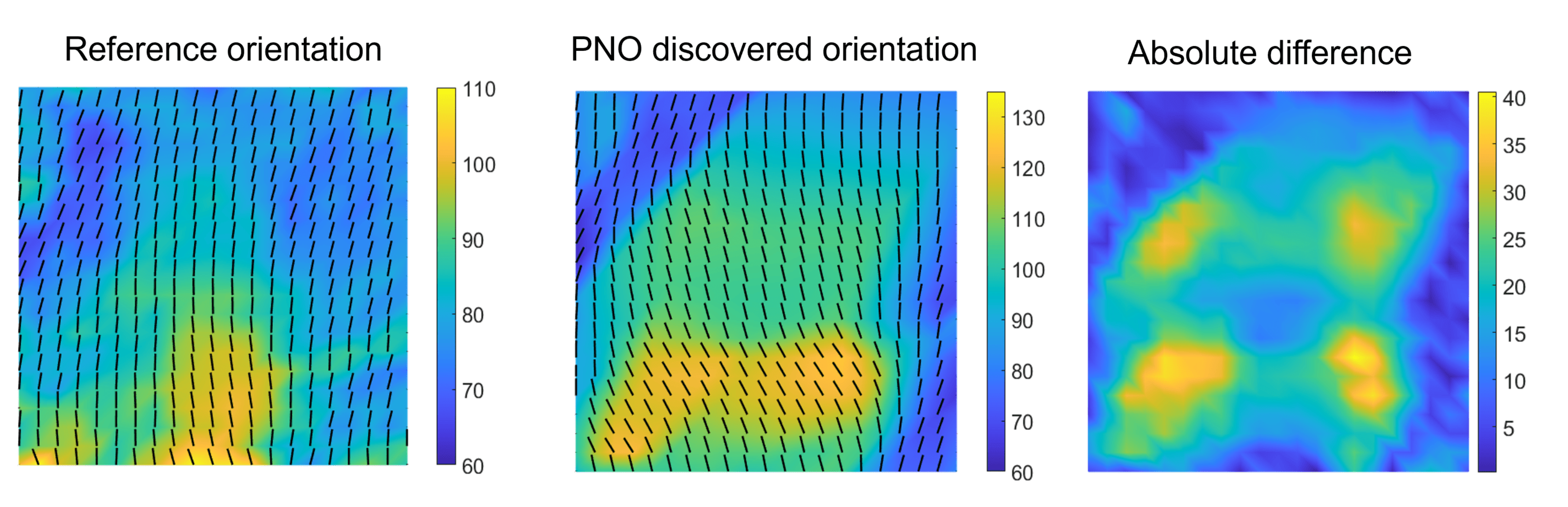}
 \caption{Prediction of collagen fiber orientation field by the hetero-PNO II model versus \YY{the experimentally detected orientation}. The averaged \YY{absolute difference is 16.0$^{\text{o}}$}.}
 \label{fig:results_fiber}
\end{figure}

\begin{figure}[!t]\centering
\includegraphics[width=.80\linewidth]{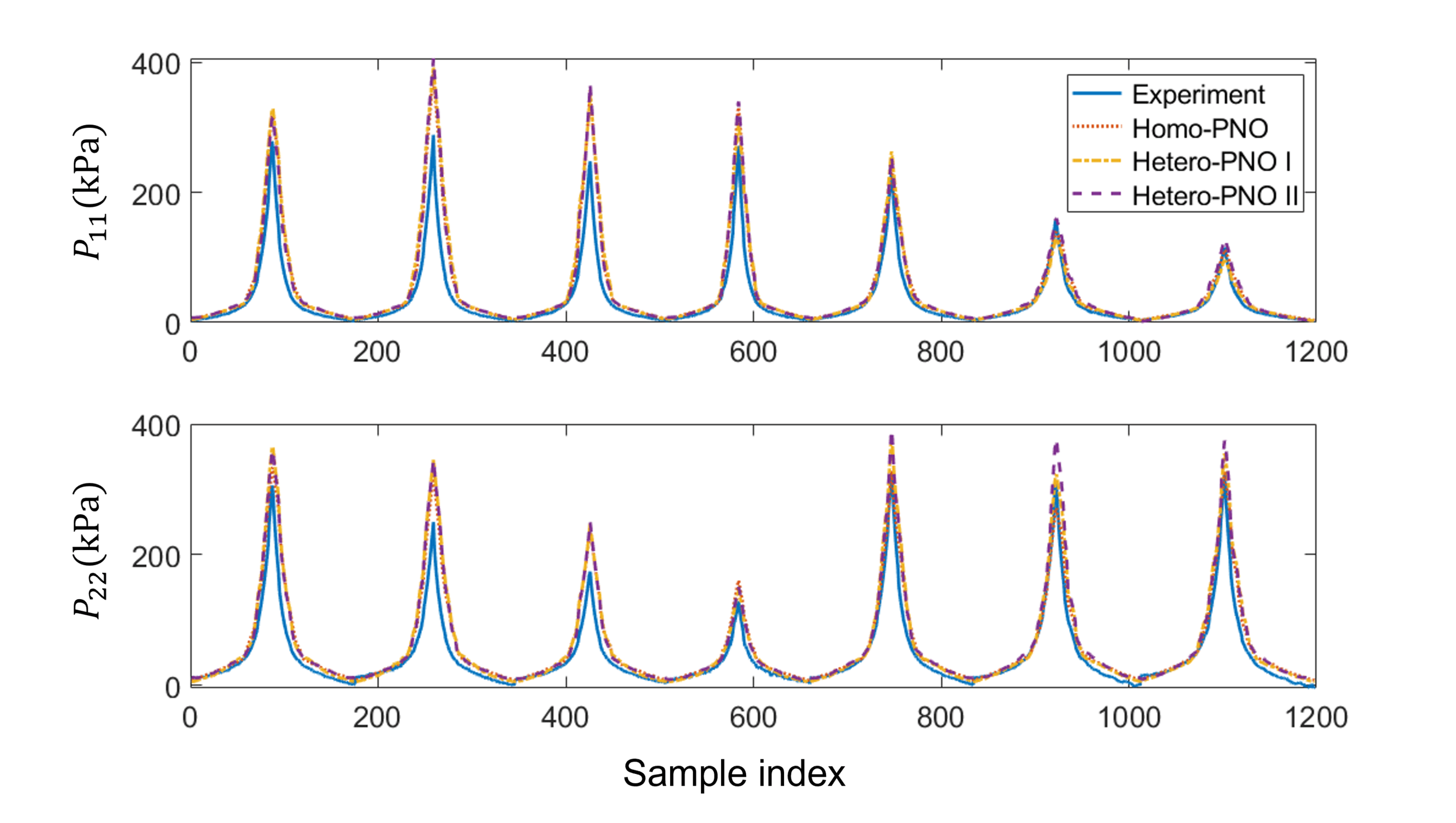}
 \caption{Predicted averaged first Piola-Kirchhoff axial stresses by the HomoPNO and HeteroPNO models versus \YY{the measured values} for the 1198 test samples.}
 \label{fig:tissue_cyclic_stress}
\end{figure}

\begin{figure}[!t]\centering
\includegraphics[width=.80\linewidth]{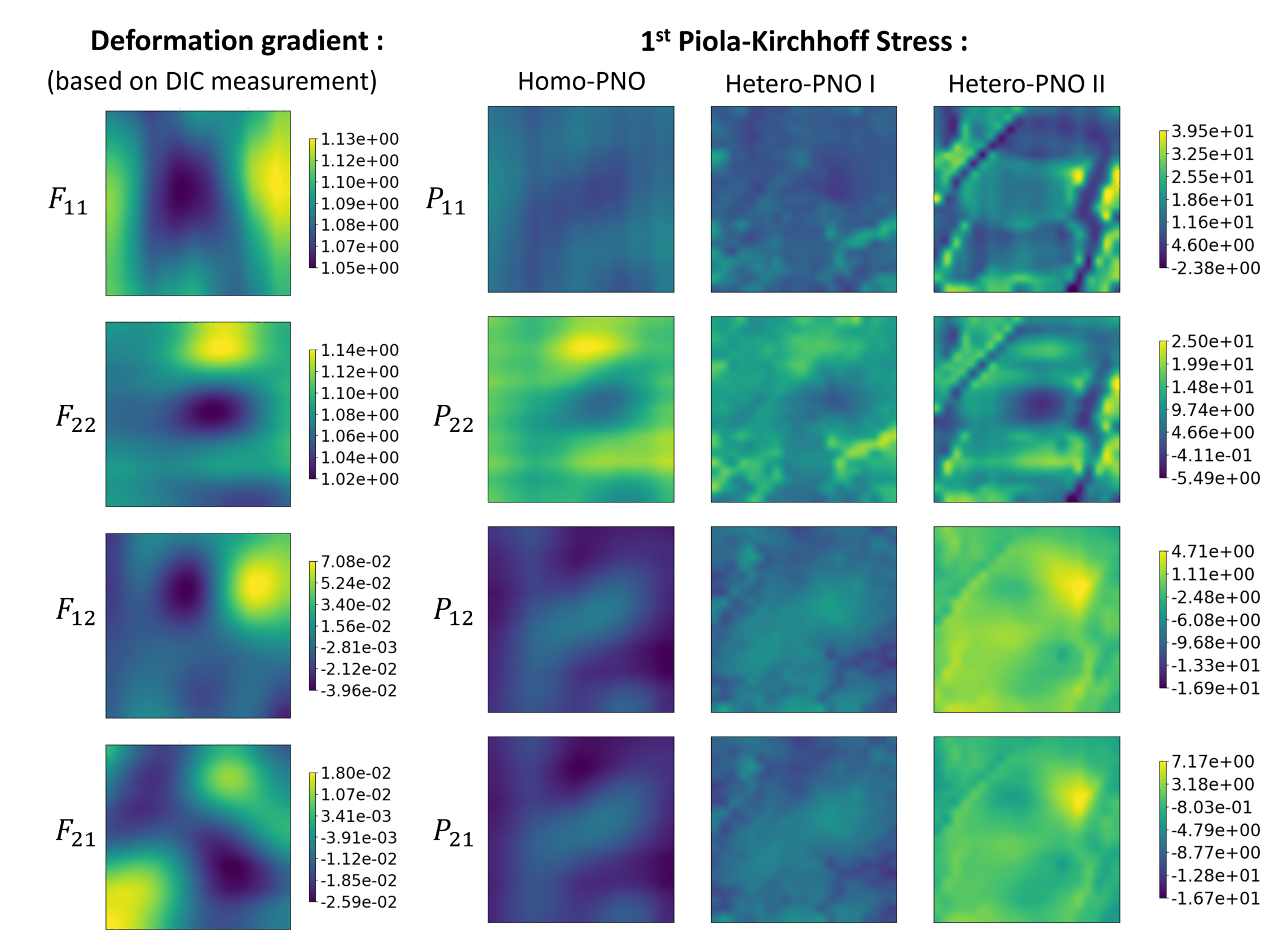}
 \caption{\YY{Predicted stress field (first Piola-Kirchhoff) by the HomoPNO and HeteroPNO models and the corresponding deformation gradient field computed from DIC-measured displacements. Stress legend values are in KPa.}}
 \label{fig:tissue_stress_field}
\end{figure}

\begin{figure}[!t]\centering
\includegraphics[width=.79\linewidth]{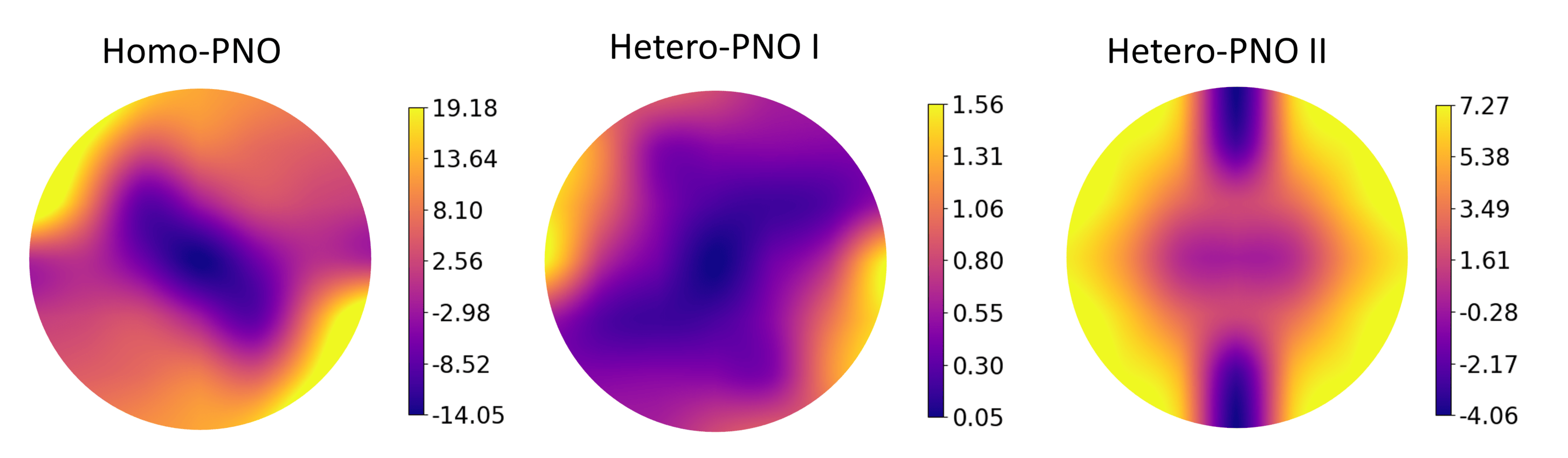}
 \caption{Learned influence states for HomoPNO and the two HeteroPNO models for the bio-tissue.}
 \label{fig:tissue_kernels}
\end{figure}

\paragraph{\textbf{Learning of homogenized and heterogeneous PNOs.}} For training our PNO models, we use a similar approach as in the previous example which is to first train a HomoPNO model, and then use the trained $\sigma^{NN}$ MLP as the initialization for $\sigma^{NN}$ in the HeteroPNO I and II. For HeteroPNO II, fiber angles are initialized as $90^{\text{o}}$, as the collagen fibers in the specimen are known to have a ``preferred direction'' of $90^{\text{o}}$ from the biology of the cardiovascular tissue where the specimen is cut out. 

In this example, since the external forces are all zero, the setting falls under the category of Case II, which minimizes the displacement error and averaged axial stress errors (see Algorithm \ref{alg:ML}). \YY{The ``FNM+$\alpha_{BC}$'' approach is employed, since pSDFI measured $\alpha$ is available on the exterior and that this approach avoids wasting portions of data for describing volume constraints in the ``VC'' approach. The loss function is then constructed as:
\begin{equation}\label{eqn: lossAlpha_caseII}
    \text{loss} = (1 - \gamma)(loss_u) + \gamma \frac{\vertii{\alpha - \alpha_{BC}}_{l^2(\tilde{\omg}_I)}}{\vertii{\alpha_{BC}}_{l^2(\tilde{\omg}_I)}} \text{ .} 
\end{equation}
where $\gamma$ is set to: 0.2, 0.5, 0.8 during hyperparameter tuning.} As mentioned in previous sections, Case II is \YY{computationally} more demanding. Therefore we use smaller MLPs for this problem: The widths of the MLPs for $\omega^{NN}$, and $\sigma^{NN}$ are respectively: (2, 32, 64, 1), and (4, 64, 64, 1), \YY{with the number of trainable parameters reported in Table \ref{table:time_memory}. We keep the width of $\alpha^{NN}$ MLP the same as before, (2, 128, 128, 1), for it to remain sufficiently expressive for capturing complex fiber patterns}. Similar to the previous example the peridynamic horizon size is set as $\delta = 3\Delta x$. In hyperparameter tuning performed during the training, $\beta$ in the loss function \YY{\eqref{eqn: caseII_loss}} is tuned with three different values: 0.2, 0.5, 0.8. \YY{In our tests, $\beta = 0.5$ for the HomoPNO, and $\beta = 0.8$ for Hetero-PNO I and II, and $\gamma = 0.5$ have lead to our best trained models. The rest of the hyperparameters are the same as those listed in Table \ref{table:HGO_hyperparam}, except for the batch size which is set to 1 for this problem. 
Table \ref{table:time_memory} provides the comparison between different PNO's when they are used with Case I or Case II loss functions depending on the problem. As observed while the Case II models have 20 times less parameters compared to the Case I models, the allocated memory is over 100 time larger and the training time is over 20 times longer. As described earlier, this is due to the iterative solver running per data instance for each epoch in the Case II setting.
}

After training the baseline (Fung model) and the PNOs, we report the displacement errors on training validation and test sets in Table \ref{table:tissue_errors}. As observed HomoPNO outperforms the baseline classical model by $35\%$, while the heterogeneous architectures improved the accuracy by another $25\%$ showing the significance of considering heterogeneity in modeling bio-tissues. \YY{
In both the homogeneous and heterogeneous cases, comparing the Fung's model with homoPNO indicates the superiority of the our data-driven approach in accuracy, compared to the expert-constructed approach.} \YY{Similar to the previous example we notice that HeteroPNO II has a slightly lower error compared to HeteroPNO I. As mentioned earlier a possible reason can the fact that HeteroPNO I is forced to work with a microstructure that contains variations in length scales smaller than $\delta$, which the PNO is not able to resolve due to its nonlocality. Another possible explanation for this example is that the reference fiber orientation field used in HeteroPNO I is based on measurement and can have some discrepancy with the actual ground-truth collagen fiber due to errors introduced by the measurement methodology, equipment, and postprocessing procedures such as smoothing. Therefore, it is possible that some areas of the learned $\alpha$ are actually closer to the ground truth compared to the map obtained from processed pSDFI measurements. This hypothesis is also consistent with the observations from Fung models:  the heterogeneous Fung model has a slightly larger error compared to the homogeneous Fung model. Among all cases, the implicit neural operator model, FNO, has the smallest error. However, we would like to highlight that the error from heterogenous PNO is of the same scale, with additional guarantees on physically interpretability and consistency.} The displacements predicted by PNOs versus the \YY{DIC measurements} for a test sample are plotted in Figure \ref{fig:tissue_disp}. The discovered microstructure via HeteroPNO II model is plotted against the experimentally measured ones in Figure \ref{fig:results_fiber}. The averaged predicted axial Piola-Kirchhoff stresses by PNO models are plotted for all 1,198 test samples against the experimentally measured values in Figure \ref{fig:tissue_cyclic_stress}. \YY{The sample indexing in this figure is according to the temporal sequence recorded during the biaxial test. One can recognize the seven protocol (loading/unloading cycles) by comparing to Figure \ref{fig:UF_biaxial}. Each spike corresponds to the maximum tension of that protocol after which the unloading starts.} As another important contribution of this work, we are able to provide the spatial stress fields for DIC displacements. Figure \ref{fig:tissue_stress_field} shows the 1st Piola-Kirchhoff stress fields obtained from PNO models for a test sample's experimental displacements, along with the deformation gradient field computed from DIC displacements via finite difference approximation. HeteroPNO stresses reveal locations with significantly higher than average stress concentrations while they are absent in the HomoPNO stress plots. These concentration sites will be susceptible to tear and rupture much earlier that the rest of the tissue, a potentially useful feature that is undetectable by homogeneous models\YY{, and difficult (if not impossible) to detect via direct measurements}. By comparing the HeteroPNO stresses and the fiber orientation field (Figure \ref{fig:fiber_exp}), it is clear that locations with rapid collagen fiber orientation transitions are susceptible to high stress concentration and possibly damage initiation.

Finally, the learned influence states of the PNO models are plotted in Figure \ref{fig:tissue_kernels}. While the preferred direction of the fibers in specimen are known to be about $90^{\text{o}}$, we see a $45^{\text{o}}$ angle in the influence state of the HomoPNO model. The reason for this phenomenon is that the DIC data only contains biaxial tension instances, which makes the data blind to $\pm 45^{\text{o}}$ anisotropy. Since the data is indifferent to existence of fibers along $\pm45^{\text{o}}$ direction, HomoPNO influence states can be affected and have larger values along $\pm45^{\text{o}}$ direction. The HeteroPNO influence states however are not affected, because they learn a fundamental influence state for the real fiber angles (or learned fiber angles that minimize the errors). As such, they show the stiffer direction along $x-$axis corresponding to $\alpha=0$ as expected by design. \YY{One may also notice that the influence states of the HeteroPNO I and HeteroPNO II cases differ. In fact, we do not anticipate to learn a unique influence state. In peridynamic modeling, kernels/influence states of different forms can be used for the same problem, where one usually choose a particular form of $\omega$ and then calibrate the constants from data or classical properties \cite{chen2015selecting, seleson2011role}. Given that $\omega$ in our study is flexible and defined via a neural network, the learned kernels can have various forms.} \YY{As shown in Figure \ref{fig:results_fiber}, HeteroPNO II learns the fiber orientation field with an averaged error of $16^\circ$. Compared to the synthetic dataset examples, the error in this example is relatively large, possibly due to the fact that the reference fiber orientation field was obtained from pSFDI measurements, which also contains noises.}


\section{Summary and future directions}\label{sec:conclusion}

In this work, we have developed a novel neural operator architecture, which we call the Heterogeneous Peridynamic Neural Operator (HeteroPNO), for modeling the mechanical response of a biological tissue specimen and discovering its fiber orientation field. By learning a nonlocal constitutive law within the framework of a heterogeneous ordinary state-based peridynamics, our HeteroPNO features 1) the guarantee of fundamental physical laws (momentum conservation laws and objectivity); 2) generalizability to different domains, loading and boundary conditions; and 3) the disentanglement of the physical effects of material nonlinear response and heterogeneity. The heart of the architecture is a neural operator with two shallow neural networks: a nonlocal kernel $\omega^{NN}$ to capture the fiber orientation, and a nonlocal model form $\sigma^{NN}$ for the peridynamic scalar force state to capture the complex constitutive behavior.

Based on the HeteroPNO architecture, a data-driven computing workflow has been proposed, which learns the material model together with fiber orientation field directly from displacement/loading data pairs. As such, we integrate material identification and microstructure inference into one unified learning framework, and the learnt model is applicable to unseen loading conditions. To verify the method on soft tissue samples with unknown spatial heterogeneity, measurement noise, fiber-reinforced and nonlinear behaviors, we successfully modeled a porcine heart TVAL specimen using DIC measurement data collected from biaxial and constrained uniaxial tension tests. By learning the material heterogeneity together with constitutive laws, our numerical studies have demonstrated improved accuracy of displacement predictions relative to alternative methods: HeteroPNO has outperformed the homogenized Fung-type model by 50\% and a homogeneous PNO model by over 25\%. The inferred microstructure and stress field show good consistency with experimental measurements. These results suggest that even in a small and noisy training data regime ($\leq100$ samples), the HeteroPNO model could build from DIC data a parametric description for the heterogeneous constitutive model  that predicts the displacement field, the microstructure, and the local stress.

Despite the encouraging results presented here, questions and potential applications require further investigations. A natural future extension is the generalization of the model to other specimens with different material properties and microstructures. In the current work, the HeteroPNO model does not account for the load-dependent reorientation and realignment of the collagen fibers \cite{fitzpatrick2022ex}. To capture these effects, one possible approach is to extend the model using transfer-learning techniques such as the meta-learning methods proposed in \cite{zhang2023meta}. Translating the current trained model to whole organ simulations would be another interesting generalization problem. \YY{We also point out that in the current work we adopt an averaged (homogenized) thickness setting, but variable thickness can also play a critical role in modeling tissue behavior \cite{lin2023impact} and therefore warrants an importants future work.} Another important future step is to accelerate the solution procedure of the learnt nonlocal constitutive model, since the computational treatment of integral equations in peridynamics is often slower than with the classical PDE solvers. To achieve speedup, we plan to investigate the combination of the neural constitutive model and the neural solution operator, possibly with an encoder/decoder architecture. \YY{Lastly, as pointed out in Section \ref{sec:algorithm}, the current PNO approach does not guarantee model well-posedness, and we rely on a validation strategy to enhance model robustness. To fully alleviate this issue, systematic design of a well-posed PNO would also be a topic for future research.}

\end{document}